\documentclass[12pt,
preprint,
superscriptaddress,
amssymb,
aps,
nofootinbib,
prd
]{revtex4-2}
\usepackage{graphicx}
\usepackage{amsmath,amsfonts}
\usepackage{mathtools}
\usepackage{slashed}
\usepackage{verbatim}
\usepackage{hyperref}
\usepackage{bm}
\usepackage{xcolor}
\usepackage{physics}
\usepackage{booktabs}
\usepackage{multirow}

\usepackage{etoolbox}
\makeatletter
\patchcmd{\frontmatter@abstract@produce}
  {\penalty-200\relax}
  {\penalty10000\relax}
  {}
  {\PackageError{abstract-layout}{REVTeX patch failed}{}}
\makeatother

\numberwithin{equation}{section}

\let\originalleft\left
\let\originalright\right
\renewcommand{\left}{\mathopen{}\originalleft}
\renewcommand{\right}{\mathopen{}\originalright}

\begin{document}
\title{ALP interactions with mesons and baryons at \texorpdfstring{$\mathcal{O}(1/f_a)$}{O(1/fa)}:\\ a general 
field redefinition invariant approach}
\author{Jorge Alda}
\email{jorge.aldagallo@unipd.it}
\affiliation{Dipartimento di Fisica e Astronomia ``Galileo Galilei'', Universit\`a di Padova \& INFN Sezione di Padova, via Marzolo 8, 35131 Padova, Italy}
\affiliation{Centro de Astropart\'iculas y F\'isica de Altas Energ\'ias (CAPA), Pedro Cerbuna 12, 50009 Zaragoza, Spain.}
\author{Stefano Rigolin}
\email{stefano.rigolin@pd.infn.it}
\affiliation{Dipartimento di Fisica e Astronomia ``Galileo Galilei'', Universit\`a di Padova \& INFN Sezione di Padova, via Marzolo 8, 35131 Padova, Italy}

\begin{abstract}
We develop a field-redefinition-invariant description of hadronic amplitudes
outside the WZW sector at $\mathcal{O}(1/f_a)$,
including general flavour-conserving and flavour-changing quark couplings.
Starting from the most general CP-conserving Lagrangian for
ALP interactions with the QCD sector, we identify
invariant Wilson-coefficient combinations and match them onto the $U(3)$ chiral
theory. Using LSZ reduction, we choose fields for which ALP--meson mixing
vanishes at this order; amplitudes then depend only on direct invariant
couplings, while the anomalous gluonic interaction remains encoded in the
$U(1)_A$ sector.
For the baryon octet, we derive the general vector and axial couplings and
identify, to our knowledge for the first time, an independent invariant
ALP--baryon interaction from a second topological singlet structure in the
$U(3)$ baryon Lagrangian. We constrain flavour-changing couplings with rare hyperon decays and
show in an explicit ultraviolet model that running and matching can generate
them from flavour-diagonal high-scale couplings. Finally, we estimate
baryon--antibaryon widths and find that, in a gluophilic benchmark, their sum
can exceed individual exclusive mesonic widths for
$2~\text{GeV} \le m_a \le 3~\text{GeV}$.
\end{abstract}

\maketitle
\clearpage
\section{Introduction}

Axion-like particles (ALPs)~\cite{Georgi:1986df,Jaeckel:2010ni} are pseudoscalar particles whose Lagrangian has an approximate 
shift symmetry, broken by anomalous gauge interactions and by small soft-breaking terms 
that provide a non-vanishing mass, $m_a$, to this pseudo-Nambu--Goldstone boson (pNGB). 
The presence of these terms spoils the mass-scale relation $m_a f_a \sim m_\pi f_\pi$ that 
characterizes the canonical QCD axion, for which the explicit breaking of the global $U(1)_{PQ}$ 
symmetry arises solely from nonperturbative QCD effects. This possibility
enlarges the phenomenologically interesting parameter space and, in particular, allows 
pseudoscalar states with masses in the MeV--GeV range, well above the typical QCD-axion 
region. In parallel, several extensions of the Peccei--Quinn (PQ) mechanism have recently 
revived the possibility of heavy QCD axions, in which the solution to the strong CP problem 
is retained while the axion mass is parametrically enhanced with respect to the canonical QCD 
axion band~\cite{Gaillard:2018xgk,Hook:2019qoh,Kivel:2022emq,Gavela:2023tzu,deGiorgi:2024elx,
Murayama:2026ioh,DiLuzio:2026poh,FernandezNavarro:2026cyu}. These developments provide 
strong motivation for a systematic study of generic ALP interactions in the hadronic, 
GeV-scale regime.

A complementary motivation arises from the flavour sector. In flaxion or axiflavon models, 
for instance, the PQ symmetry is identified with a Froggatt--Nielsen flavour symmetry, linking 
the strong CP and flavour problems and naturally giving rise to non-universal and potentially 
flavour-changing axion couplings to fermions~\cite{Ema:2016ops,Calibbi:2016hwq,deLaVega:2021ugs}.
More generally, the independence of $m_a$ and $f_a$ in ALP theories, together with the possibility 
of nontrivial flavour structures, makes the MeV--GeV mass range particularly rich from both
the theoretical and phenomenological perspectives.

Of particular interest in the MeV--GeV mass range are the interactions of axions and ALPs with hadrons. 
At sufficiently low energies, $m_a \lesssim 1\,\mathrm{GeV}$, these interactions can be systematically 
described within Chiral Perturbation Theory ($\chi$PT), while for $m_a \gtrsim 2\!-\!3\,\mathrm{GeV}$ 
inclusive hadronic observables can be addressed using perturbative QCD (pQCD). 
The intermediate region is considerably more challenging. There, the chiral expansion progressively
loses its predictive power, while the perturbative description has not yet set in, and a variety of 
hadronic resonances beyond the light pseudoscalar nonet can give sizeable contributions to ALP production
and decay amplitudes. A consistent description of this transition region is phenomenologically important, as
the total hadronic width determines the ALP lifetime and hence whether its decay occurs promptly, with
a displaced vertex, or outside the detector, while the exclusive branching fractions and production rates 
determine the experimental signatures and sensitivities of searches for GeV-scale 
ALPs~\cite{Aloni:2018vki,Alda:2025uwo}.

A first systematic attempt to bridge this gap was provided in Ref.~\cite{Aloni:2018vki}, where the low-energy chiral description was supplemented by a data-driven treatment 
of hadronic resonances and form factors, allowing exclusive ALP decay rates to be 
estimated up to masses of a few GeV. However, the implementation followed in 
Refs.~\cite{Aloni:2018vki,Jerhot:2022pbg,DallaValleGarcia:2024gvi} leads to theoretical 
inconsistencies, as the calculated amplitudes retain an unphysical dependence on the 
arbitrary ALP-dependent chiral rotation used to remove the ALP--gluon interaction.

This issue has recently been addressed from complementary perspectives in Refs.~\cite{Ovchynnikov:2025gpx,
Balkin:2025enj}. In Ref.~\cite{Ovchynnikov:2025gpx}, invariance under field redefinitions is restored by consistently 
including both the (direct) ALP--hadron interactions generated by the chiral rotation and the (indirect) 
contributions induced by ALP--meson mixing, whose basis-dependent pieces cancel in physical amplitudes. 
In Ref.~\cite{Balkin:2025enj}, basis-independent amplitudes are obtained
by combining direct and ALP--meson-mixing contributions into an effective
ALP matrix. Its invariance is demonstrated from the transformation
properties of its constituents, although its explicit expression is built
from mixing parameters that are individually basis dependent; a general
closed expression solely in terms of invariant couplings is not provided.
Both approaches also address the extension of the low-energy description
towards the resonance region. In particular,
Ref.~\cite{Ovchynnikov:2025gpx} enlarges the hadronic spectrum by explicitly
including heavier pseudoscalar and axial-vector excitations, in addition to
the scalar, vector and tensor resonances already employed in data-driven
descriptions, while Ref.~\cite{Balkin:2025enj} incorporates resonance
contributions and hadronic form factors to extend the chiral amplitudes
towards the few-GeV regime. These developments emphasise that, once the ALP
mass approaches the hadronic scale, a description restricted to the light
pseudoscalar nonet is no longer sufficient and additional hadronic degrees
of freedom must be taken into account. The odd intrinsic parity 
Wess--Zumino--Witten (WZW) sector, which is not considered here, has been treated 
separately in field redefinition invariant formulations for two and three 
light flavours in Refs.~\cite{Bai:2024lpq,Bai:2025fvl}.

In these recent developments, the existing basis-independent
formulations have so far been restricted to flavour-conserving ALP--quark
interactions. This assumption simplifies the structure of the
effective theory, but is not generic from an ultraviolet (UV) model-building
model building. Non-universal flavour structures arise naturally in several
well-motivated scenarios and, after rotating to the quark mass basis and
accounting for renormalisation-group evolution and matching, may induce flavour-changing
ALP interactions even when they are absent at the UV scale.

Moreover, the fact that such interactions are strongly constrained by rare meson decays
does not provide, by itself, a reason to discard them from a general low-energy description.
Indeed, flavour-changing processes
may constitute some of the most sensitive probes of the underlying ALP couplings. Restricting the
hadronic theory to flavour-conserving interactions would therefore limit
its ability to match and test generic UV models. A fully general treatment
requires extending the field redefinition invariant construction to
non-diagonal flavour structures and identifying the corresponding invariant
combinations that enter physical amplitudes.

In this work, we develop a general field-redefinition-invariant framework
for ALP interactions with hadrons at $\mathcal{O}(1/f_a)$, without assuming
flavour-diagonal ALP--quark couplings. In Sec.~\ref{sec:QCDLag}, we start
from the most general CP-conserving ALP--QCD Lagrangian at this order and
study its transformation under general axial and vector quark field
redefinitions. This allows us to identify the combinations of Wilson
coefficients that are invariant under such transformations, including
those associated with flavour-changing interactions. 

In Sec.~\ref{sec:chiral}, we match these invariant couplings onto the
low-energy $U(3)$ chiral theory in the Callan--Coleman--Wess--Zumino (CCWZ)
formalism. For amplitudes with
one external ALP outside the odd intrinsic parity WZW sector, we use Lehmann--Symanzik--Zimmermann (LSZ)
reduction to identify the contributions induced by ALP--meson mixing.
We then exploit the freedom of ALP-dependent quark field redefinitions
to choose a parametrisation in which these mixing contributions vanish
at $\mathcal{O}(1/f_a)$. The amplitudes can consequently be obtained from
direct vertices written solely in terms of field redefinition invariant
combinations, while the effects of the partonic ALP--gluon coupling are
retained through the $U(1)_A$ topological sector.

As a first application of this framework beyond the mesonic sector, in
Sec.~\ref{sec:baryons} we derive the most general interactions of the ALP
with the baryon octet at $\mathcal{O}(1/f_a)$, including both
flavour-conserving and flavour-changing couplings. In doing so, we retain a
second singlet axial structure, originally introduced in the general
$U(3)$ baryon chiral Lagrangian of Refs.~\cite{Bass:1999is,Borasoy:1999nd,Borasoy:2000xj} to account for
the external QCD vacuum angle. This structure was subsequently discarded
there because the external topological source was irrelevant to the
processes under consideration. When the topological source contains a
dynamical ALP, however, this structure generates an independent invariant
ALP--baryon interaction proportional to the anomalous ALP--gluon coupling.
To our knowledge, the resulting ALP--baryon interaction and its phenomenological
consequences are studied here for the first time. This observation may also 
be relevant for astrophysical and cosmological constraints, some of which 
depend on the effective ALP couplings to protons and neutrons~\cite{Lella:2022uwi,
Caputo:2024oqc,Lella:2024dmx}. 

The resulting effective couplings are used in
Sec.~\ref{sec:lightpheno} to investigate ALP phenomenology in baryonic
processes. We first study flavour-changing hyperon decays into ALPs, using
the recent results for
$\Sigma^+\to p+\mathrm{inv}$ and
$\Xi^0\to\Lambda+\mathrm{inv}$ from
BESIII~\cite{BESIII:2023utd,BESIII:2026ztz}, and for $\Sigma^+\to p\mu^+\mu^-$ from
LHCb~\cite{LHCb:2017rdd,LHCb:2025evf}, to constrain the corresponding
ALP--baryon couplings. As an explicit example of the matching to a
well-defined UV model, we then consider the QED--DFSZ model of
Ref.~\cite{Alda:2025uwo}. Using the \texttt{ALP-aca}
package~\cite{Alda:2025nsz}, we evolve its flavour-diagonal high-scale
couplings to low energies and compare the constraints from mesonic and
baryonic flavour-changing processes associated with the $s\to d a$
transition. This explicit example illustrates why flavour-changing interactions 
must be retained even when the UV boundary conditions are flavour diagonal.
Finally, in Sec.~\ref{sec:baryonpairs} we investigate ALP decays into
baryon--antibaryon pairs in the few-GeV region. We construct a
phenomenological continuation of the low-energy ALP--baryon couplings in
terms of baryonic form factors and use it to estimate the exclusive partial
widths above the corresponding kinematic thresholds. We present our
conclusions in Sec.~\ref{sec:conclusions}.

\section{The effective ALP-QCD  Lagrangian at \texorpdfstring{$\mathcal{O}(1/f_a)$}{O(1/fa)}}
\label{sec:QCDLag}

The most general CP-conserving Lagrangian describing the ALP interactions with gluons and $N_f$ quark flavours,
up to $\mathcal{O}(1/f_a)$, can be written as
\begin{equation}\begin{split}
\mathcal{L} =& -\frac{1}{4}G_{\mu\nu}^a G^{\mu\nu,a} +\overline{q}\left(i\slashed{D}-\bm{m_q}\right)q
+ \frac{1}{2}(\partial_\mu a) (\partial^\mu a) - \frac{1}{2} m_a^2 \,a^2 - V(a) \\
& - i\frac{a}{f_a}\overline{q}\big([ \bm{c_q^S}, \bm{m_q}]-\{\bm{c_q^P},\bm{m_q}\}\gamma_5)q + 
 \frac{\partial_\mu a}{f_a}\overline{q}\gamma^\mu(\bm{c_q^V} + \bm{c_q^A}\gamma_5)q + 
\frac{\alpha_s  c_G}{4\pi}\frac{a}{f_a}G_{\mu\nu}^a \widetilde{G}^{\mu\nu,a}.
\label{eq:lagr_partons}
\end{split}\end{equation}

In Eq.~\eqref{eq:lagr_partons}, $q$ denotes an $N_f$-component column
vector in flavour space, $\bm{m_q}$ is the diagonal and real
$N_f\times N_f$ quark mass matrix, and
$\bm{c_q^S}$, $\bm{c_q^P}$, $\bm{c_q^V}$, and $\bm{c_q^A}$
are Hermitian $N_f\times N_f$ matrices, 
not necessarily diagonal, when quarks with identical electric charge are included in the same flavour multiplet, 
i.e. for $N_f>2$. The potential $V(a)$ contains ALP self-interaction terms that will not be considered further in 
this work. Interactions of the ALP with leptons and electroweak gauge bosons are also omitted, since our focus is 
on the ALP--hadron sector.

It is well known that the parametrisation of the Lagrangian in Eq.~\eqref{eq:lagr_partons} is redundant, as
physically equivalent descriptions are related by ALP-dependent quark field redefinitions.
A generic ALP-dependent quark chiral field redefinition, $\bm{g}=(\bm{g_L},\bm{g_R})\in U(N_f)_L \times U(N_f)_R$, 
can be written as
\begin{equation}
 q_{L,R} \to \bm{g_{L,R}} \,q_{L,R} \equiv 
      \exp\left(\frac{i a}{f_a} (\bm{\delta} \mp \bm{\kappa})\right) q_{L,R} , \label{eq:chitranf}
\end{equation}
where $\bm{\delta}$ and $\bm{\kappa}$ are $N_f\times N_f$ Hermitian matrices, again, not necessarily diagonal 
when $N_f>2$ is assumed. Under this field redefinition the ALP couplings in Eq.~\eqref{eq:lagr_partons} 
are shifted as:
\begin{equation}\begin{split}
    c_G \to c_G - \langle \bm{\kappa}\rangle,\qquad  \bm{c_q^{P}} \to 
    \bm{c_q^P} + \bm{\kappa},\qquad \bm{c_q^A} \to \bm{c_q^A} + \bm{\kappa},\\
    \bm{c_q^S} \to \bm{c_q^S} +\bm{\delta},\qquad \bm{c_q^V} \to 
    \bm{c_q^V}+\bm{\delta}.
\end{split}
\label{eq:shifts}
\end{equation}
Here and in the following, $\langle \cdots \rangle \equiv \operatorname{tr}(\cdots)$ is understood. The parameters 
in the Hermitian matrices $\bm{\kappa}$ and $\bm{\delta}$ connect physically equivalent descriptions of the 
ALP-hadronic sector. It follows from Eq.~\eqref{eq:shifts} that the combinations of couplings 
\begin{equation}\begin{split}
    c_G + \langle \bm{c_q^A}\rangle \equiv c_{GA},&\qquad \bm{c_q^A} - \bm{c_q^P}\equiv \bm{c_q^{AP}},\\
    c_G + \langle \bm{c_q^P}\rangle\equiv c_{GP}, &\qquad \bm{c_q^V} -\bm{c_q^S} \equiv \bm{c_q^{VS}},
\end{split}\label{eq:invariants}
\end{equation}
are invariant under the field redefinitions $\bm{g_{L,R}}$. Eq.~\eqref{eq:invariants} therefore provides a convenient
overcomplete set of field-redefinition-invariant combinations of Wilson coefficients at $\mathcal{O}(1/f_a)$, 
in terms of which all physical observables can be expressed.

The Lagrangian in Eq.~\eqref{eq:lagr_partons} is more general than those commonly considered in related studies,
such as Refs.~\cite{Aloni:2018vki,Ovchynnikov:2025gpx,Balkin:2025enj}.
Several comments are therefore in order.

First, the redundancies in the parametrisation of Eq.~\eqref{eq:lagr_partons} should be interpreted with care. 
If only QCD and QED interactions are considered, a diagonal vector redefinition of the quark fields can be used 
to remove the diagonal entries of $\bm{c_q^V}$. The same redefinition shifts $\bm{c_q^S}$, whose diagonal entries, 
however, do not contribute to the scalar interaction, since $[\bm{c_q^S},\bm{m_q}]$ vanishes for diagonal $\bm{c_q^S}$. 
Hence, diagonal scalar couplings are unphysical, while the flavour off-diagonal information in the vector--scalar 
sector is encoded in the invariant combination $\bm{c_q^{VS}}=\bm{c_q^V}-\bm{c_q^S}$. In particular, when 
the theory is restricted to flavour-diagonal couplings and to QCD and QED interactions, the entire vector--scalar sector can be 
removed by a suitable choice of field variables. 

This conclusion no longer holds once weak flavour-changing interactions are included, since electroweak (EW) 
interactions do not preserve the individual quark flavour symmetries~\cite{Bauer:2021wjo}. A local flavour-dependent 
vector redefinition that removes a diagonal entry of $\bm{c_q^V}$ induces ALP-dependent phases in charged-current
interactions, which below the electroweak scale are encoded in $\Delta F=1$ weak operators. Although we do not 
explicitly perform the EW matching in this work, these considerations motivate retaining the vector--scalar sector 
when aiming for a complete description of the low-energy ALP phenomenology.

More importantly for the applications considered below, flavour off-diagonal ALP couplings cannot in general
be consistently neglected. Even if the ALP couplings are flavour diagonal at a scale $\Lambda\simeq4\pi f_a > 
v_{\rm ew}$, SM running and matching generate flavour-changing couplings at low energies. In particular, diagrams 
involving the top quark and $W^\pm$ boson generate flavour violating contributions to the left-handed down-quarks  
combination $\bm{c_q^L}=\bm{c_q^V}-\bm{c_q^A}$. A framework in which flavour-violating derivative couplings are omitted 
from the outset, as in Ref.~\cite{Ovchynnikov:2025gpx,Balkin:2025enj}, does not capture this generic low-energy flavour structure. 
Since these couplings mediate rare meson decays which typically provide the strongest constraints~\cite{Bauer:2021mvw,Alda:2025uwo}, omitting 
them can overstate the viable parameter space. We therefore retain general flavour structures in all ALP--quark 
coupling matrices from the outset.

Furthermore, even when ALP flavour-violating couplings are allowed, only the shift-symmetric ALP--fermion couplings,
$\bm{c_q^V}$ and $\bm{c_q^A}$, are generally introduced in the Lagrangian. This choice reflects the standard 
axion-like assumption that the ALP interactions with fermions respect the shift symmetry, with its breaking 
arising only through the ALP potential and anomalous gauge interactions. In some UV constructions, however, scalar and pseudoscalar ALP couplings, 
$\bm{c_q^S}$ and $\bm{c_q^P}$, may also be present. They may encode additional sources of explicit
shift-symmetry breaking in the UV theory, as occurs for instance in the Majoron scenario of 
Ref.~\cite{deGiorgi:2024elx}, or arise from field redefinitions in theories 
with a spontaneously broken global $U(1)_{\rm PQ}$ symmetry, such as DFSZ models~\cite{Zhitnitsky:1980tq,Dine:1981rt}.

Indeed, even if one chooses a basis in which $\bm{c_q^S}=\bm{c_q^P}=0$, an ALP-dependent field redefinition
of the form given in Eq.~\eqref{eq:chitranf} leads to an equivalent parametrisation in which the quark mass matrix becomes ALP-dependent,
\begin{equation}
\bm{m_q}\to \bm{m_q} (a) = e^{i\frac{a}{f_a} (\bm{\delta} - \bm{\kappa})} \, \bm{m_q} \, 
e^{-i\frac{a}{f_a} (\bm{\delta} + \bm{\kappa})}=\bm{m_q} -i\frac{a}{f_a} \{\bm{m_q}, \bm{\kappa}\} - 
i\frac{a}{f_a} [\bm{m_q},\bm{\delta}] + \mathcal{O}(1/f_a^{2}). \nonumber
\end{equation}
As an example, consider the chiral rotation introduced in Ref.~\cite{Georgi:1986df} to remove the anomalous 
ALP--gluon coupling. In our parametrisation, its effect at $\mathcal{O}(1/f_a)$ is automatically encoded in the 
shift of the pseudoscalar coupling, $\bm{c_q^P}\to\bm{c_q^P}+\bm{\kappa}$, with  $\langle\bm{\kappa}\rangle=c_G$.
We keep a constant quark mass matrix
$\bm{m_q}$ and encode the corresponding ALP dependence in explicit scalar
and pseudoscalar ALP--quark interactions. This is equivalent at
$\mathcal{O}(1/f_a)$ to the ALP-dependent mass matrix parametrisation used
in Ref.~\cite{Balkin:2025enj}.
Similarly, in the presence of flavour 
off-diagonal ALP couplings, an axial field redefinition satisfying $[\bm{c_q^A},\bm{\kappa}]\neq0$ generates, 
through the derivative ALP--quark interaction, a tower of higher-order operators of the form $a^n\partial_\mu a\,
(\bar q\gamma^\mu\gamma_5^n q)/f_a^{n+1}$. These higher-order terms are consistently beyond the accuracy 
considered here, since throughout we truncate the effective field theory (EFT) at $\mathcal{O}(1/f_a)$.

The main advantage of the parametrisation in
Eq.~\eqref{eq:lagr_partons} is that it makes equivalence under ALP-dependent
quark field redefinitions manifest while allowing us to work from the outset
with a Lagrangian consistently truncated at a fixed order in $1/f_a$.
It also makes the connection among different $U(N_f)$ chiral
parametrisations transparent and allows the matching onto the low-energy
hadronic EFT to be performed directly in an arbitrary quark field
parametrisation. In particular, the matching does not require a preliminary
chiral rotation removing the anomalous ALP--gluon coupling. This feature will
become explicit in the construction of the hadronic theory presented in the
following section.

\section{Chiral realisation and invariant amplitudes at \texorpdfstring{$\mathcal{O}(1/f_a)$}{O(1/fa)}}
\label{sec:chiral}

The field redefinition invariants identified in
Sec.~\ref{sec:QCDLag} provide the quark-level sources for the low-energy
chiral theory. Their CCWZ~\cite{Coleman:1969sm,Callan:1969sn} realisation determines the mesonic interactions
and the quadratic ALP--meson system. We use LSZ reduction to identify the
mesonic components of the physical ALP and then exploit the freedom of
quark field redefinitions to choose a parametrisation in which these
components vanish at $\mathcal{O}(1/f_a)$. In this parametrisation,
one-ALP amplitudes outside the WZW
sector\footnote{Throughout this work, ``non-anomalous'' refers specifically
to amplitudes outside the odd intrinsic parity WZW sector. The anomalous
$U(1)_A$ singlet sector is nevertheless included because it encodes the
QCD topological effects required to match the ALP--gluon interaction.}
are obtained from direct vertices with manifestly invariant couplings,
while the ALP--gluon interaction is matched through the anomalous
$U(1)_A$ sector.

\subsection{CCWZ embedding}

To construct the low-energy theory, we use the nonlinear realisation of
$G=U(N_f)_L\times U(N_f)_R$, spontaneously broken by the quark condensate
to the diagonal subgroup $H=U(N_f)_V$. We collect the pseudoscalar fields in
\begin{equation}
 \bm{\Phi} \equiv \sum^{N^2_f-1}_{i=0}\bm{\lambda_i}\Phi_i \underset{N_f=3}{=}
\begin{pmatrix}
\pi_3 + \frac{\eta_8}{\sqrt{3}} + \sqrt{\frac{2}{3}}\eta_0 & \sqrt{2}\,\pi^+ & \sqrt{2}\,K^+ \\
\sqrt{2}\,\pi^- & -\pi_3 + \frac{\eta_8}{\sqrt{3}} + \sqrt{\frac{2}{3}}\eta_0 & \sqrt{2}\,K^0 \\
\sqrt{2}\,K^- & \sqrt{2}\,\bar{K}^0 & -\frac{2\eta_8}{\sqrt{3}} + \sqrt{\frac{2}{3}}\eta_0
\end{pmatrix},
\label{eq:pseudoscalar}
\end{equation}
where the $U(N_f)$ generators satisfy $\langle\bm{\lambda_i}\bm{\lambda_j^\dagger}\rangle=2\delta_{ij}$. The fields $\Phi_i$ are defined 
in the flavour generator basis and do not yet coincide with the physical neutral meson mass eigenstates\footnote{The off-diagonal generators correspond to charge eigenstates and are therefore non-Hermitian. For charged pions, for example, $\pi^-=(\pi^+)^\dagger$ and $\bm{\lambda_{\pi^-}}=\bm{\lambda_{\pi^+}}^\dagger$.}. ALP-induced mixing is treated perturbatively for $f_\pi/f_a\ll1$, while App.~\ref{app:mesonmass} describes the separate numerical determination of the neutral SM mass eigenstates.

Below $\Lambda_\mathrm{QCD}$, the appropriate degrees of freedom are hadrons. The CCWZ construction~\cite{Coleman:1969sm,Callan:1969sn} realises $G\to H$ non-linearly and provides common building blocks for mesons, baryons, and heavier resonances. We use it here to match the partonic interactions in Eq.~\eqref{eq:lagr_partons} onto the hadronic theory while keeping their field-redefinition properties explicit.

We parametrise the coset $G/H$ by the representative
\begin{equation}
    \bm{u} = \exp\left(\frac{i}{2f_\pi} \bm{\Phi}\right) 
\end{equation}
with $f_\pi=93\,\mathrm{MeV}$. Under the ALP-dependent quark field redefinition in Eq.~\eqref{eq:chitranf},
\begin{equation}
    \bm{u} \to \bm{g_R}\,\bm{u}\,\bm{h}^\dagger = \bm{h}\,\bm{u}\,\bm{g_L}^\dagger,\label{eq:chiraltransf_mesons}
\end{equation}
where $\bm h\in U(N_f)_V$ is the CCWZ compensator. At $\mathcal{O}(1/f_a)$, the compensator reads
\begin{equation}\begin{split}
\bm{h} & = \bm{1}+ \frac{a}{f_a}\left(i\bm{\delta} + \tan\left(\frac{\mathbb{C}_{\bm{\Phi}}}{4f_\pi}\right)[\bm{\kappa}] \ \right) \\ & = \bm{1}+ \frac{a}{f_a}\left(i\bm{\delta} +\frac{1}{4f_\pi}[\bm{\Phi},\bm{\kappa}]+\frac{1}{192 f_\pi^3}[\bm{\Phi},[\bm{\Phi},[\bm{\Phi}, \bm{\kappa}]]] + \ldots\right),
\end{split}\end{equation}
with the linear operator $\mathbb{C}_{\bm{\Phi}}[\bm{\kappa}] \equiv [\bm{\Phi},\bm{\kappa}]$.  
Accordingly, under a general ALP-dependent quark field redefinition at $\mathcal{O}(1/f_a)$,
the field $\bm{u}$ and the pseudoscalar mesons nonet transform as:
\begin{equation}\begin{split}
\bm{u} \to {}&\bm{u} + i\frac{a}{f_a}\left(\bm{\kappa}\bm{u} + [\bm{\delta},\bm{u}] +i \bm{u} \tan\left(\frac{\mathbb{C}_{\bm{\Phi}}}{4f_\pi}\right)[\bm{\kappa}] \right)\, \\ 
\bm\Phi\to {}&\bm\Phi+\frac{a}{f_a}
\left\{
i[\bm\delta,\bm\Phi]
+
\left[
\mathbb C_{\bm\Phi}
\cot\left(\frac{\mathbb C_{\bm\Phi}}{2f_\pi}\right)
\right][\bm\kappa]
\right\}\\
={}&\bm\Phi+\frac{a}{f_a}\bigg[
i[\bm\delta,\bm\Phi]
+2f_\pi\bm\kappa
-\frac{1}{6f_\pi}
[\bm\Phi,[\bm\Phi,\bm\kappa]]+\cdots\bigg].\label{eq:Phi_transformation_resummed}
\end{split}\end{equation}
In particular, the singlet transformation law reads
$\langle\bm\Phi\rangle \to \langle\bm\Phi\rangle + \frac{2 f_\pi}{f_a}a\langle\bm\kappa\rangle$.
The derivation of the previous formula is detailed in App.~\ref{sec:Phi_expansion}.

In the presence of external ALP vector and axial currents coupled to quarks, the CCWZ vielbein, namely 
the axial-vector component of the Maurer-Cartan one-form, reads:
\begin{equation}
\bm{u_\mu} \equiv i\big(\bm{u}^\dagger\partial_\mu\bm{u}-\bm{u}\partial_\mu \bm{u}^\dagger \big) + 
\frac{\partial_\mu a}{f_a}[\bm{u}^\dagger(\bm{c_q^V} + 
\bm{c_q^A})\bm{u}-\bm{u}(\bm{c_q^V}-\bm{c_q^A})\bm{u}^\dagger].
\label{eq:vielbein}
\end{equation}
It follows directly from this definition that $\bm{u_\mu}$ transforms linearly under the ALP-dependent chiral 
transformations, i.e. $\bm{u_\mu}\to\bm{h}\,\bm{u_\mu}\,\bm{h}^\dagger$. At $\mathcal{O}(1/f_a)$, this covariance 
realises in the CCWZ construction the axial and vector source transformations in Eq.~\eqref{eq:shifts}.

For any field $\bm{X}$ transforming linearly under chiral transformations, i.e. $\bm{X}\to \bm{h} \bm{X} 
\bm{h}^\dagger$, the CCWZ covariant derivative $\bm{\nabla_\mu} \bm{X} = \partial_\mu \bm{X} + 
[\bm{\Gamma_\mu},\bm{X}]$ is defined in terms of the ALP-dependent covariant connection, namely the vector 
component of the Maurer--Cartan one-form, as follows.
\begin{eqnarray}
    \bm{\Gamma_\mu} \equiv \frac{1}{2}\big(\bm{u}^\dagger\partial_\mu\bm{u}+\bm{u}\partial_\mu \bm{u}^\dagger \big)-i\frac{\partial_\mu a}{2f_a}[\bm{u}^\dagger(\bm{c_q^V} + \bm{c_q^A})\bm{u}+\bm{u}(\bm{c_q^V}-\bm{c_q^A})\bm{u}^\dagger].
\label{eq:connection}
\end{eqnarray}
Neglecting $\mathcal{O}(1/f_a^2)$ terms, $\bm{\nabla_\mu}\bm{X}$ transforms covariantly. Expanding the building 
blocks in powers of the pNGB field $\bm{\Phi}/f_\pi$, gives, up to quadratic order in fields,
\begin{equation}\begin{split}
    \bm{u_\mu} &= 2\frac{\partial_\mu a}{f_a}\bm{c_q^A} -\frac{1}{f_\pi}\partial_\mu \bm{\Phi} 
    -i\frac{\partial_\mu a}{f_\pi f_a}[\bm{\Phi},\bm{c_q^V}] + \mathrm{terms~with~3~or~more~fields},\\
    \bm{\Gamma_\mu} &= -i \frac{\partial_\mu a}{f_a}\bm{c_q^V}+\frac{1}{8f_\pi^2}[\bm{\Phi},\partial_\mu\bm{\Phi}]-\frac{\partial_\mu a}{2 f_\pi f_a}[\bm{\Phi},\bm{c_q^A}]+
    \mathrm{terms~with~3~or~more~fields}.
\end{split}\end{equation}
The derivation of these results and the corresponding all-orders expansions in $\bm{\Phi}$ at $\mathcal{O}(1/f_a)$ 
can be found in Appendix~\ref{sec:Phi_expansion}.

Quark masses and the scalar and pseudoscalar ALP interactions enter in the chiral Lagrangian through the 
symmetry-breaking spurion
\begin{equation}
\bm{\chi} \equiv 2B_0\big(\bm{m_q}+i\frac{a}{f_a}[\bm{c_q^S},\bm{m_q}]+i\frac{a}{f_a}\{\bm{c_q^P},\bm{m_q}\}\big),
\label{eq:massbreaking}
\end{equation}
whose first term is the usual source of pseudoscalar masses. The corresponding combinations that transform covariantly 
under the unbroken subgroup $H$ are
\begin{equation}
\bm{\chi_\pm} = \bm{u}^\dagger \bm{\chi}\bm{u}^\dagger \pm \bm{u} \bm{\chi}^\dagger\bm{u}.
\label{eq:massterm}
\end{equation}
To quadratic order in the fields, one has
\begin{equation}\begin{split}
    \bm{\chi_+} & = 4 B_0 \left( \bm{m_q} + \frac{1}{2 f_\pi}\{\bm{\Phi},\{\bm{c_q^P},\bm{m_q}\}\}
    \frac{a}{f_a}-\frac{1}{8 f_\pi^2} \{\bm{\Phi},\{\bm{\Phi},\bm{m_q}\}\} \right) \\ 
    & + 4 i B_0\,[\bm{c_q^S},\bm{m_q}]\frac{a}{f_a} + \mathrm{terms~with~3~or~more~fields},
\end{split}
\end{equation}

Eqs.~\eqref{eq:vielbein}, \eqref{eq:connection}, and \eqref{eq:massbreaking} provide the low-energy counterparts 
of the axial, vector, scalar, and pseudoscalar ALP-quark interactions. Since these building blocks transform covariantly, 
the matching can be performed directly from an arbitrary quark field parametrisation.

\subsection{The \texorpdfstring{$U(1)_A$}{U(1)A} sector}
\label{sec:U1A}

The gluonic source in Eq.~\eqref{eq:lagr_partons} is incorporated into the chiral theory through the anomalous 
$U(1)_A$ sector of the $U(3)$ construction~\cite{Herrera-Siklody:1996tqr,Kaiser:2000gs}. At leading order (LO) in the chiral and large-$N_c$ expansions, this sector is represented by the
auxiliary topological-charge density
\begin{equation}
\omega = \frac{\alpha_s}{8\pi}G_{\mu\nu}^a \widetilde{G}^{\mu\nu,a} ,\label{eq:topo_density}
\end{equation}
the leading chiral and large-$N_c$ Lagrangian that reproduces the anomaly and topological susceptibility 
$\chi_t$ is~\cite{DiVecchia:1980yfw}
\begin{equation}
    \mathcal{L}_{TC} = \frac{\omega^2}{2\chi_t} - 2 i\, \omega \left(\langle\log \bm{u}\rangle + 
    c_G \frac{i\,a}{f_a}\right).\label{eq:X_U1A}
\end{equation}
The equation of motion identifies the field-redefinition-invariant topological source
\begin{equation}
     X_{TC} \equiv\frac{\omega}{\chi_t} = 2\,i\left(\langle\log\bm{u}\rangle + c_G\frac{i\,a}{f_a}\right)\, .
\label{eq:topological}
\end{equation}
Integrating out $\omega$ yields the Witten--Veneziano term~\cite{Witten:1979vv,Veneziano:1979ec},
\begin{equation}
    \mathcal{L}_{TC} = 2 \chi_t \bigg(\langle\log\bm{u}\rangle + c_G \frac{i\,a}{f_a} \bigg)^2 = 
                     - \frac{f_\pi^2 m_0^2}{ 4N_f}X_{TC}^2 ,
\end{equation}
where $m_0^2\equiv2N_f\chi_t/f_\pi^2$. The contribution of the gluonic coupling to covariant operators 
is encoded only through $X_{TC}$ and its derivatives. Combining $\partial_\mu\langle\log\bm u\rangle
=\langle\bm u^\dagger\partial_\mu\bm u\rangle$ with Eq.~\eqref{eq:vielbein} gives
\begin{equation}
\partial_\mu X_{TC} = \langle\bm{u_\mu}\rangle - 
2\frac{\partial_\mu a}{f_a}c_{GA} = 
-\frac{1}{f_\pi}\partial_\mu\langle\bm{\Phi}\rangle-2c_G \frac{\partial_\mu a}{f_a} .
\label{eq:dX_U1A}
\end{equation}
Thus, $\partial_\mu X_{TC}$ provides an independent covariant singlet building block for the chiral effective theory, and it will play a crucial role in the construction of the baryon Lagrangian.

\subsection{Quadratic meson Lagrangian and the SM mass basis}

The chiral Lagrangian can be organised according to the simultaneous
expansion in derivatives, light-quark masses, and $1/N_c$ introduced
in Refs.~\cite{Leutwyler:1996sa,Kaiser:2000gs}, with
\begin{equation}
    p=\mathcal{O}(\sqrt{\epsilon}) \, , \qquad
    \bm{m_q} \sim \mathcal{O}(\epsilon) , \qquad
    1/N_c=\mathcal{O}(\epsilon) .
\end{equation}
In this counting, the standard chiral quantities scale as
\begin{equation}
\bm{u}=\mathcal{O}(\epsilon^0),\qquad
\bm{u_\mu},\bm{\Gamma_\mu}=\mathcal{O}(\sqrt{\epsilon}),
\qquad
\bm{\chi_\pm}=\mathcal{O}(\epsilon),\qquad
f_\pi=\mathcal{O}(1/\sqrt{\epsilon}).
\end{equation}
We further assign
\begin{equation}
X_{TC}=\mathcal{O}(\epsilon^0),\qquad
\frac{a}{f_a}(c_G,\bm{c_q})=\mathcal{O}(\epsilon^0).
\end{equation}

At LO in the simultaneous chiral and large-$N_c$ expansion,
and retaining ALP interactions up to $\mathcal{O}(1/f_a)$, the mesonic
Lagrangian is given by
\begin{equation}\begin{split}
\mathcal{L} & = \frac{f_\pi^2}{4}\langle \bm{u_\mu}\bm{u^\mu} + \bm{\chi_+}\rangle 
    +\frac{1}{2}\partial_\mu a \partial^\mu a - \frac{1}{2}m_{a,0}^2 a^2 
    - \frac{f_\pi^2 m_0^2}{ 4N_f}X_{TC}^2 \\
    &\supset \frac{1}{2}\left(\partial_\mu \Phi_I\right)^\dagger K_{IJ}\, \left(\partial^\mu \Phi_J\right) - 
    \frac{1}{2}\Phi_I^\dagger M_{IJ}^2 \Phi_J.
    \label{eq:PGBeftLO}
\end{split}\end{equation}
Terms such as $X_{TC}^k$ with $k\geq3$ or $(\partial_\mu X_{TC})(\partial^\mu X_{TC})$ are suppressed in the large $N_c$ counting and first enter at next-to-leading order (NLO) in the $\epsilon$ expansion.

In Eq.~\eqref{eq:PGBeftLO}, the uppercase index
$I=(i,\hat a)$ runs over the $N_f^2$ pseudoscalar fields and the ALP.
The lowercase index $i=0,\ldots,N_f^2-1$ labels the SM pseudoscalar
fields, while $\hat a$ denotes the ALP, $\Phi_{\hat a}=a$.
At $\mathcal{O}(1/f_a)$, the Hermitian kinetic and mass matrices
$K_{IJ}$ and $M_{IJ}^2$ are given by
\begin{equation}
\begin{array}{lcl}
    K_{\hat{a}\hat{a}} = 1 + \mathcal{O}(1/f_a^{2}) , &\qquad &
    M^2_{\hat{a}\hat{a}} = m_{a,0}^2 + \mathcal{O}(1/f_a^{2}), \\[10pt]
    K_{\hat{a}i} = K_{i\hat{a}}^\dagger = -\dfrac{f_\pi}{f_a}\langle\bm{c_q^A}\bm{\lambda_i}\rangle , & \qquad &
    M_{\hat{a}i}^2 = (M_{i\hat{a}}^2)^\dagger = -\dfrac{f_\pi}{f_a}\left( B_0\langle\{\bm{c_q^P},\bm{m_q}\} \bm{\lambda_i}\rangle - 
    \dfrac{m_0^2}{N_f} c_G \langle\bm{\lambda_i}\rangle \right), \\[10pt]
    K_{ij} = \dfrac{1}{2}\langle \bm{\lambda_i}^\dagger\bm{\lambda_j}\rangle = \delta_{ij}, & \qquad &
    M^2_{ij}=\dfrac{B_0}{2} \langle\bm{m_q}\{\bm{\lambda_i}^\dagger,\bm{\lambda_j}\}\rangle + 
    \dfrac{m_0^2}{2N_f}\langle\bm{\lambda_i^\dagger}\rangle\langle\bm{\lambda_j}\rangle . 
\end{array}
\label{eq:KandM}
\end{equation}

The neutral fields $\pi_3$, $\eta_8$, and $\eta_0$ in the flavour
generator basis are not mass eigenstates. Even in the absence of the
ALP, they mix to form the physical mesons $\pi^0$, $\eta$, and $\eta'$.
ALP interactions further perturb this system, but the resulting shifts
of the SM meson masses start at $\mathcal{O}(1/f_a^2)$ and can therefore
be neglected at the order considered here. We first diagonalise the SM
meson sector and subsequently treat ALP mixing as a small 
$\mathcal{O}(f_\pi/f_a)$ perturbation, assuming $f_\pi \ll f_a$.
The numerical diagonalisation is described in App.~\ref{app:mesonmass}.

Henceforth, the index $i$ labels physical pseudoscalar mesons, and 
$\bm{\lambda_{\pi^0}}$, $\bm{\lambda_\eta}$, and
$\bm{\lambda_{\eta'}}$ denote the associated algebra generators. 
In this basis, the SM mass matrix is diagonal,
\begin{equation}
M_{ij}^2=m_{i,0}^2\delta_{ij}+\mathcal{O}(1/f_a^2),
\end{equation}
with $m_{i,0}=m_i$ at $\mathcal{O}(1/f_a)$.

\subsection{LSZ treatment of ALP--meson mixing}

We now use the LSZ reduction formula to account for ALP--meson mixing 
in amplitudes at order $\mathcal{O}(1/f_a)$. At this order, the mixing effects can be 
included through meson pole contributions without explicitly diagonalising the full ALP--meson
quadratic Lagrangian. For non-canonical quadratic terms, LSZ associates the field $\Phi_J$ that interpolates 
a physical particle $\mathfrak p$ of mass $m_{\mathfrak p}$ with an eigenvector $w_J^{(\mathfrak p)}$ 
satisfying
\begin{equation}
    (K_{IJ} m_\mathfrak{p}^2 - M^2_{IJ})w_J^{(\mathfrak{p})} = 0, \qquad w_I^{(\mathfrak{p}),*} K_{IJ} w_J^{(\mathfrak{q})} = \delta_{\mathfrak{p} \mathfrak{q}}.
\end{equation}
For the ALP, $\mathfrak p=a$, this condition becomes
\begin{equation}
    (K_{i\hat{a}}m_a^2-M^2_{i\hat{a}})w_{\hat{a}}^{(a)} + (m_a^2-m_{i}^2)w_i^{(a)}=0 \, ,
\end{equation}
with $m_{a,0}=m_a$ at $\mathcal{O}(1/f_a)$. The coefficients $w_i^{(a)}$ therefore measure the pseudoscalar-meson
components of the physical ALP state.
Since $w_i^{(a)}=\mathcal{O}(f_\pi/f_a)$ and $w_{\hat a}^{(a)}=\mathcal{O}(1)$, we set $w_{\hat a}^{(a)}=1$. Neglecting $\mathcal{O}(f_\pi^2/f_a^2)$ terms, we obtain 
\begin{equation} 
    w_I^{(a)} = \begin{cases}
        \displaystyle 1 & I=\hat{a}\\
        \displaystyle \frac{f_\pi}{f_a}\left(\frac{\langle\bm{\mathcal{C}}\bm{\lambda_i}^\dagger\rangle}{m_{i}^2-m_a^2} + \langle\bm{c_q^A}\bm{\lambda_i}^\dagger\rangle\right) & I = i
    \end{cases} 
    \label{eq:mixmesons}
\end{equation}
with the field-redefinition invariant Hermitian matrix
\begin{equation}
    \bm{\mathcal{C}} \equiv -B_0 \{\bm{m_q},\bm{c_q^{AP}}\}-\frac{m_0^2}{N_f}c_{GA}\bm{1}. \label{eq:mixing_coeff}
\end{equation}

At $\mathcal{O}(1/f_a)$, the LSZ-reduced amplitude
receives two contributions. The first is the
``direct'' contribution, obtained from interaction vertices containing an
explicit ALP field. The second is the ``mixing'' contribution, arising from
the mesonic components $w_i^{(a)}$ of the physical ALP state and the
corresponding amplitudes with an external physical pseudoscalar meson. 
The separation between direct and mixing contributions is, however, parametrisation dependent. In particular, the coefficients $w_i^{(a)}$ are not invariant under ALP-dependent quark field redefinitions, since they specify the decomposition of the physical ALP state for a given choice of field variables.

This freedom can be exploited to choose a parametrisation in which the physical ALP has no mesonic component at $\mathcal{O}(1/f_a)$, namely $w_i^{(a)}=0$ for all physical pseudoscalar states. To express this choice compactly, we define the linear operator
\begin{equation}
\mathbb{M}_a(\bm A)
\equiv
B_0\{\bm{m_q},\bm A\}
+\frac{m_0^2}{N_f}\langle\bm A\rangle\bm 1
-m_a^2\bm A.
\label{eq:Ma_operator}
\end{equation}
In the SM mass basis, its inverse admits the spectral representation
\begin{equation}
\mathbb{M}_a^{-1}(\bm A)
=
\frac{1}{2}\sum_i
\frac{\langle\bm A \bm{\lambda_i}^\dagger\rangle}
{m_i^2-m_a^2}\,
\bm{\lambda_i},
\label{eq:Ma_inverse}
\end{equation}
where the sum runs over the complete set of physical pseudoscalar
states. A detailed derivation and an explicit expression suitable for
practical calculations are given in App.~\ref{sec:app_completeness}
(see Eq.~\eqref{eq:Ma_inverse_full}).

Under an ALP-dependent field redefinition the coefficients $w_i^{(a)}$ transform as
\begin{equation} 
w_i^{(a)}\to w_i^{(a)}+\frac{f_\pi}{f_a}\langle\bm\kappa\bm{\lambda_i}^\dagger\rangle.
\end{equation}
The choice of field redefinition 
\begin{equation}
\bm{\kappa} = -\frac{f_a}{2 f_\pi} \sum_i w_i^{(a)} \bm{\lambda_i} =
            -\bm{c_q^A}-\mathbb M_a^{-1}(\bm{\mathcal C}).
\label{eq:kappa_LSZ}
\end{equation}
therefore removes the ALP--meson mixing components completely at $\mathcal{O}(1/f_a)$.
Indeed, after this redefinition the mixing coefficients become
\begin{equation}
\widehat{w}_i^{(a)}
=
\frac{f_\pi}{f_a}
\left[
\frac{\langle\bm{\mathcal C}\bm{\lambda_i}^\dagger\rangle}
{m_i^2-m_a^2}
+
\langle(\bm{c_q^A}+\bm\kappa)\bm{\lambda_i}^\dagger\rangle
\right]
=0.
\label{eq:wi_LSZ_zero}
\end{equation}
Therefore, in this parametrisation, all amplitudes at order $1/f_a$ can be obtained entirely from the corresponding direct interactions.

At the level of the quadratic ALP--meson Lagrangian, only the  term in Eq.~\eqref{eq:Phi_transformation_resummed}  that is independent of $\bm\Phi$ contributes to the mixing coefficients. With the choice in Eq.~\eqref{eq:kappa_LSZ}, the relevant transformation is
\begin{equation}
    \bm{\Phi} \to  \bm{\Phi} - 2 \frac{f_\pi}{f_a}\big( \bm{c_q^A} + \mathbb{M}_a^{-1}(\bm{\mathcal{C}})\big)a,
\label{eq:shifted_mesons}
\end{equation}

The same field redefinition shifts the pseudoscalar field matrix 
and reshuffles the ALP--quark and ALP--gluon couplings according to Eq.~\eqref{eq:shifts}. With the choice in Eq.~\eqref{eq:kappa_LSZ}, they become
\begin{equation}
\begin{split}
    \widehat{c}_G
    &=
    c_{GA}
    +\left\langle\mathbb M_a^{-1}(\bm{\mathcal C})\right\rangle,
    \\
    \bm{\widehat{c}_q^P}
    &=
    -\bm{c_q^{AP}}
    -\mathbb M_a^{-1}(\bm{\mathcal C}),
    \\
    \bm{\widehat{c}_q^A}
    &=
    -\mathbb M_a^{-1}(\bm{\mathcal C}).
\end{split}
\label{eq:ALPcouplings_nomixing}
\end{equation}
These expressions generalise the transformations of Refs.~\cite{Ovchynnikov:2025gpx,Balkin:2025enj} to non-diagonal flavour structures. Since $\bm{\mathcal C}$ is field-redefinition invariant, every quantity on the right-hand side of Eq.~\eqref{eq:ALPcouplings_nomixing} is expressed solely in terms of the invariant combinations introduced in Eq.~\eqref{eq:invariants}. Thus, although the condition $w_i^{(a)}=0$ selects a particular parametrisation, the resulting direct ALP couplings are uniquely determined by field-redefinition-invariant quantities. Physical amplitudes computed in this parametrisation are therefore manifestly independent of the quark field variables used in the original ALP--QCD Lagrangian.

The behaviour of this parametrisation becomes particularly transparent in
the light and heavy ALP limits. For $m_a\ll m_\pi$, retaining the complete
$\eta_0$ contribution at finite $m_0$, the inverse operator reduces to
\begin{equation}
    \mathbb M_a^{-1}(\bm{\mathcal C})
    \;\longrightarrow\;
    -\bm{c_q^{AP}}
    -c_{GP}\,
    \frac{\bm{m_q}^{-1}}
    {\langle\bm{m_q}^{-1}\rangle+2N_f B_0/m_0^2}.
\label{eq:Ma_light_limit}
\end{equation}
In the large $m_0$ limit, corresponding to the decoupling of the
$\eta_0$, this expression becomes
\begin{equation}
    \mathbb M_a^{-1}(\bm{\mathcal C})
    \;\longrightarrow\;
    -\bm {c_q^{AP}}
    -c_{GP}\,
    \frac{\bm{m_q}^{-1}}
    {\langle\bm{m_q}^{-1}\rangle}.
\label{eq:Ma_light_large_m0}
\end{equation}
The latter is the mass dependent expression commonly adopted in the ALP
literature to remove ALP--meson mixing
\cite{Georgi:1986df,Aloni:2018vki,Ovchynnikov:2025gpx}.
For $m_a\gg m_{\eta'}$ one has
\begin{equation}
    \mathbb M_a^{-1}(\bm{\mathcal C})
    \;\longrightarrow\;
    -\frac{\bm{\mathcal C}}{m_a^2}.
\label{eq:Ma_heavy_limit}
\end{equation}

This construction also clarifies the relation to previous treatments of
ALP--meson mixing. In the conventional approach, a specific
ALP-dependent quark field redefinition is performed before matching onto
the chiral theory, and physical amplitudes are obtained by combining
parametrisation dependent direct and mixing contributions. The recent
basis-independent treatments of
Refs.~\cite{Ovchynnikov:2025gpx,Balkin:2025enj} demonstrate explicitly
that this dependence cancels in physical amplitudes.
Ref.~\cite{Ovchynnikov:2025gpx} implements the cancellation by combining
the direct interactions generated by the chiral rotation with the
corresponding ALP--meson mixing contributions, while
Ref.~\cite{Balkin:2025enj} packages the two contributions into an
invariant effective ALP matrix. In the latter construction, the effective
matrix is invariant, although its explicit representation is given in
terms of mixing parameters that are individually parametrisation
dependent.

Here we use LSZ reduction to determine an ALP-dependent quark field
redefinition for which the mesonic components of the physical ALP vanish
at $\mathcal{O}(1/f_a)$. The mixing contribution is then absent, and its
effects are transferred entirely to the direct vertices. The transformed
direct couplings are given in closed form solely in terms of the invariant
combinations in Eq.~\eqref{eq:invariants}. Field redefinition invariance
is therefore manifest before constructing the amplitude, and the
prescription extends directly to general non-diagonal flavour structures.

\subsection{Practical prescription and domain of validity}

The construction developed above applies to amplitudes outside the WZW
sector involving one external ALP, consistently evaluated at
$\mathcal{O}(1/f_a)$ and at LO in the simultaneous chiral and large
$N_c$ expansion. The perturbative treatment further requires
$f_\pi/f_a\ll1$ and breaks down near an ALP--meson degeneracy. More
precisely, for each physical pseudoscalar state $i$, the mixing
contribution remains perturbative provided
\begin{equation}
    |m_a^2-m_i^2|
    \gg
    \frac{f_\pi}{f_a}
    \left|
    \langle\bm{\mathcal C}\bm{\lambda_i}^\dagger\rangle
    \right| \approx \frac{f_\pi}{f_a} m_i^2.
\label{eq:mixing_validity}
\end{equation}
At exact degeneracy, and provided the corresponding off-diagonal mixing
does not vanish, the ALP and meson states become maximally mixed,
$|w_i^{(a)}|=1/\sqrt{2}$~\cite{DiLuzio:2024jip}, and the perturbative
treatment adopted here is no longer valid.
Within this domain, an amplitude outside the WZW sector with one external ALP can be calculated as follows.
\begin{enumerate}
\item Specify the effective ALP--QCD Lagrangian at the hadronic matching scale in any convenient quark field parametrisation, keeping general flavour structures in the ALP--quark couplings.

\item Construct the corresponding $U(3)$ hadronic Lagrangian at the desired order in $p$, $\bm{m_q}$, and $1/N_c$,
     using the CCWZ building blocks defined above. In the resulting quadratic Lagrangian, determine the SM pseudoscalar mass basis.

 \item Construct the operator $\mathbb M_a$ and the invariant matrix
$\bm{\mathcal C}$ entering the ALP--meson mixing coefficients. Then
implement the ALP-dependent quark field redefinition of
Eq.~\eqref{eq:kappa_LSZ}, which sets $w_i^{(a)}=0$ for all physical
pseudoscalar states at $\mathcal{O}(1/f_a)$. For the vector
transformation, one may choose $\bm\delta=-\bm{c_q^S}$, so that
$\bm{\widehat{c}_q^V}=\bm{c_q^{VS}}$ and
$\bm{\widehat{c}_q^S}=0$.

\item Expand the transformed Lagrangian to the required number of pseudoscalar fields, following App.~\ref{sec:Phi_expansion}, and compute the amplitude from direct one-ALP vertices. Every coupling is then manifestly invariant.

\end{enumerate}

This prescription can be applied starting from any quark field parametrisation of the ALP--QCD Lagrangian and leads to the same invariant direct couplings. In this parametrisation, amplitudes with one external ALP are obtained entirely from direct ALP interactions, with no separate ALP--meson mixing contribution. It therefore provides a straightforward connection between the field-redefinition invariants identified at the quark level and the computation of physical one-ALP amplitudes in the hadronic theory.

\section{ALP interactions with the baryon octet at \texorpdfstring{$\mathcal{O}(1/f_a)$}{O(1/fa)}}
\label{sec:baryons}

We now apply the prescription developed in Sec.~\ref{sec:chiral} to
derive the interactions of ALPs with the baryon octet at order $\mathcal{O}(1/f_a)$. In the
parametrisation in which the ALP--meson mixing contribution vanishes at this order, the result is encoded in direct vector and axial
ALP--baryon couplings expressed entirely in terms of
field-redefinition-invariant combinations. The construction applies to
both flavour-conserving and flavour-changing interactions at LO in the simultaneous chiral and large-$N_c$ expansion.

With $N_f=3$ flavours, the lightest spin $1/2$ baryons are collected in the octet
\begin{equation}
    \bm{\mathcal{B}} = \begin{pmatrix}
        \frac{1}{\sqrt{2}}\Sigma^0 + \frac{1}{\sqrt{6}}\Lambda & \Sigma^+ & p\\
        \Sigma^- & \frac{-1}{\sqrt{2}}\Sigma^0 + \frac{1}{\sqrt{6}}\Lambda & n\\
        \Xi^- & \Xi^0 & -\frac{2}{\sqrt{6}}\Lambda
    \end{pmatrix} = \sum_i \bm{\lambda_{B_i}} B_i.
\end{equation}
The octet transforms covariantly under the preserved subgroup $H$ of chiral transformations, i.e. $\bm{\mathcal{B}} \to \bm{h} \bm{\mathcal{B}} \bm{h}^\dagger$. As for the pseudoscalar mesons, we use a basis of (non-Hermitian) $U(N_f)$ generators 
    $\bm{\lambda_i}$ normalised as $\langle \bm{\lambda_i}
\bm{\lambda_j^\dagger}\rangle = \delta_{ij}$.

The $\mathcal{O}(p)$ chiral Lagrangian describing pseudoscalar meson--baryon interactions, including the $\eta_0$ contributions, is~\cite{Jenkins:1995gc,Borasoy:1999nd}
\begin{equation}\begin{split}
\mathcal{L}_B =& \langle \bar{\bm{\mathcal{B}}} \gamma^\mu \, i \bm{\nabla_\mu} \bm{\mathcal{B}}\rangle - 
  M_B \langle\bar{\bm{\mathcal{B}}} \bm{\mathcal{B}}\rangle + 
  \frac{D_B}{2}\langle\bar{\bm{\mathcal{B}}} \gamma^\mu \gamma_5\{\bm{u_\mu}, \bm{\mathcal{B}}\}\rangle \\ 
& +\frac{F_B}{2}\langle\bar{\bm{\mathcal{B}}} \gamma^\mu \gamma_5[\bm{u_\mu}, \bm{\mathcal{B}}]\rangle +
  \frac{S_B}{2}\langle\bar{\bm{\mathcal{B}}}\gamma^\mu\gamma_5\bm{\mathcal{B}}\rangle\langle\bm{u_\mu}\rangle , 
\label{eq:baryoncoeff}
\end{split}\end{equation}
where $M_B$ is the mass of the baryon octet in the chiral limit. See also \cite{Scherer:2012xha} for a general 
review on the construction of the baryon Lagrangian in chiral perturbation theory.
The ALP dependence is incorporated into Eq.~\eqref{eq:baryoncoeff} through the ALP 
modified vielbein $\bm{u_\mu}$ and connection $\bm{\Gamma_\mu}$, defined in Eqs.~\eqref{eq:vielbein} and 
\eqref{eq:connection}. 

These modifications, however, do not exhaust the allowed ALP-dependent structures. 
The most general $U(3)$ baryon Lagrangian at linear order in the derivative expansion contains, in addition 
to the singlet axial interaction proportional to $\langle\bm{u_\mu}\rangle$, a second independent singlet 
structure involving the covariant derivative of the external vacuum angle~\cite{Bass:1999is,Borasoy:1999nd,Borasoy:2000xj}. This term was subsequently disregarded 
because the external topological source could be set to zero for the processes under consideration in Ref.~\cite{Borasoy:1999nd}.\footnote{In the notation of Ref.~\cite{Borasoy:1999nd}, $\nabla_\mu\theta = \langle \bm{u_\mu}\rangle-\partial_\mu X_{TC}$, and therefore the coefficients are $U_4 = (S_B + \widetilde{S}_B)/2$ and $U_5=-\widetilde{S}_B/2$. }
In the present theory, the corresponding interaction is written in terms of the invariant topological
source $X_{TC}$ and must therefore be retained, as follows.
\begin{equation}
\delta\mathcal{L}_B = \frac{\widetilde S_B}{2}
\langle\bar{\bm{\mathcal B}}\gamma^\mu\gamma_5\bm{\mathcal B}\rangle \,\partial_\mu X_{TC}.
\label{eq:newsinglet}
\end{equation}

To our knowledge, the present
work is the first to retain this second structure as an independent
interaction of a dynamical ALP with the baryon octet and to study its
phenomenological consequences.

To extract the ALP interactions with two baryons at order $\mathcal{O}(1/f_a)$, we use the parametrisation defined by
Eq.~\eqref{eq:kappa_LSZ}, in which the ALP--meson mixing contribution vanishes at $\mathcal{O}(1/f_a)$. We also fix 
the remaining vector field-redefinition freedom by choosing $\bm{\delta}=-\bm{c_q^S}$, so that
$\bm{\widehat{c}_q^V}=\bm{c_q^{VS}}$ and $\bm{\widehat{c}_q^S}=0$. In this parametrisation, the relevant
components of the baryonic building blocks are
\begin{equation}
\begin{aligned}
\left.\bm{\nabla_\mu}\bm{\mathcal B}\right|_a &=
-i\frac{\partial_\mu a}{f_a}  [\bm{c_q^{VS}},\bm{\mathcal B}],
\\
\left.\bm{u_\mu}\right|_a &=
-2\frac{\partial_\mu a}{f_a} \mathbb M_a^{-1}(\bm{\mathcal C}),
\\
\left.\partial_\mu X_{TC}\right|_a &=
-2\frac{\partial_\mu a}{f_a} \left[ c_{GA}
 +\left\langle\mathbb M_a^{-1}(\bm{\mathcal C})\right\rangle \right].
\end{aligned}
\label{eq:baryon_ALP_building_blocks}
\end{equation}
Here, $\left.\cdots\right|_a$ denotes the terms linear in the ALP field and containing no pseudoscalar mesons. 

Substituting Eq.~\eqref{eq:baryon_ALP_building_blocks} into Eqs.~\eqref{eq:baryoncoeff} and 
\eqref{eq:newsinglet} gives the interaction of the ALP with the baryon octet at the chosen order
\begin{equation}
\mathcal L_{aB} =
\frac{\partial_\mu a}{f_a} \sum_{B_1,B_2}
\bar B_1\gamma^\mu \left(c_{B_1B_2}^V+c_{B_1B_2}^A\gamma_5\right)B_2,
\label{eq:baryoncoupl_full}
\end{equation}
where
\begin{equation}
\begin{aligned}
c_{B_1B_2}^V ={}&
-\left\langle \bm{c_q^{VS}} [\bm{\lambda_{B_1}}^\dagger,\bm{\lambda_{B_2}}] \right\rangle ,
\\[2mm]
c_{B_1B_2}^A ={}& 
-(D_B+F_B) \left\langle \bm{\lambda_{B_1}}^\dagger \mathbb M_a^{-1}(\bm{\mathcal C})
  \bm{\lambda_{B_2}} \right\rangle
-(D_B-F_B) \left\langle \bm{\lambda_{B_1}}^\dagger \bm{\lambda_{B_2}} 
  \mathbb M_a^{-1}(\bm{\mathcal C}) \right\rangle
\\
&- \left[ (S_B+\widetilde S_B) \left\langle\mathbb M_a^{-1}(\bm{\mathcal C})\right\rangle
+\widetilde S_B\,c_{GA} \right] \left\langle \bm{\lambda_{B_1}}^\dagger\bm{\lambda_{B_2}}\right\rangle .
\end{aligned}
\label{eq:baryoncouplings}
\end{equation}
All ALP couplings appearing in Eq.~\eqref{eq:baryoncouplings} enter through the field redefinition 
invariant combinations $\bm{c_q^{VS}}$, $\bm{\mathcal C}$, and $c_{GA}$. The effective baryon couplings are
therefore independent of the quark field parametrisation from which the matching is performed. The 
condition of vanishing ALP--meson mixing is only a convenient choice of field variables used in their derivation.

To make contact with the existing literature, we recast the covariant
Lagrangian of Eqs.~\eqref{eq:baryoncoeff} and \eqref{eq:newsinglet} in
the Heavy Baryon Chiral Perturbation Theory (HBChPT)
formalism~\cite{Jenkins:1990jv}. 
The baryon 4-momentum is split as $p_\mu = M_B v_\mu + q_\mu$, where $v^\mu$ is the 4-vector velocity of the non-relativistic baryon, satisfying the normalisation condition $v^2=1$, and $q^\mu$ is the residual small 4-momentum component. For a nearly on-shell baryon,
$q\cdot v \ll M_B, \Lambda_\mathrm{QCD}$, there is a well-defined expansion in powers of $q/M_B$ and $q/\Lambda_\mathrm{QCD}$. The baryon fields $\bm{\mathcal{B}}$ can therefore be split into the light ($\bm{\mathcal{B}_v}$) and heavy ($\bm{\mathcal{H}_v}$) $v$-dependent components. After integrating out the heavy component, the non-relativistic Lagrangian for $\bm{\mathcal{B}_v}$ reads:
\begin{equation}\begin{split}
    \mathcal{L}_{\bm{\mathcal{B}_v}} =& \langle\bm{\bar{\mathcal{B}}_v}  i v\cdot\bm{\nabla} \bm{\mathcal{B}_v}\rangle  + 
  D_B\langle\bm{\bar{\mathcal{B}}_v} S^\mu \{\bm{u_\mu}, \bm{\mathcal{B}_v}\}\rangle 
+F_B\langle\bm{\bar{\mathcal{B}}_v} S^\mu [\bm{u_\mu}, \bm{\mathcal{B}_v}]\rangle \\&+
  S_B\langle\bm{\bar{\mathcal{B}}_v} S^\mu\bm{\mathcal{B}_v}\rangle\langle\bm{u_\mu}\rangle +
  \widetilde{S}_B\langle\bm{\bar{\mathcal{B}}_v} S^\mu\bm{\mathcal{B}_v}\rangle\,\partial_\mu X_{TC},
  \label{eq:baryoncoeffNR}
\end{split}\end{equation}
    in terms of the Pauli--Lubanski spin operator $S^\mu =\frac{i}{2}\gamma_5\sigma^{\mu\nu}v_\nu$. Adapting the counting scheme of Ref.~\cite{Jenkins:1990jv} to our notation gives the additional power-counting assignments for the heavy-baryon EFT: $\bm{\mathcal{B}_v} = \mathcal{O}(\epsilon^0)$, $v_\mu = \mathcal{O}(\epsilon^0)$, and $S^\mu = \mathcal{O}(\epsilon^0)$.
    We use the heavy-baryon reduction only to establish a consistent chiral power counting for the baryon sector. All couplings derived below, including Eq.~\eqref{eq:baryoncoupl_full}, are obtained from the covariant $U(3)$ baryon Lagrangian and do not rely on a non-relativistic expansion of the amplitudes.
    The LO terms in the baryon Lagrangians of Eqs.~\eqref{eq:baryoncoeff} and~\eqref{eq:baryoncoeffNR} then appear at $\mathcal{O}(\sqrt{\epsilon})$.

It is customary to relate the low-energy constants of B$\chi$PT to the proton quark axial charges $\Delta q$ defined as~\cite{GrillidiCortona:2015jxo}
\begin{equation}
    \langle p | \bar{q} \gamma^\mu \gamma_5 q|p\rangle = s^\mu \Delta q,
\end{equation}
where $|p\rangle$ denotes the proton state at rest and $s^\mu = \langle p|\bar{p}_v S^\mu p_v|p\rangle$ is its spin-polarisation vector.
At LO, matching the B$\chi$PT Lagrangian to the QCD matrix elements gives
\begin{equation}\begin{split}
 \Delta u & = D_B + F_B + S_B,\\
 \Delta d & = S_B,\\ 
 \Delta s & =  D_B - F_B  +S_B.\\ 
\end{split} \label{eq:axial_matching}
\end{equation}
Therefore the axial charges only determine $D_B, F_B,$ and $S_B$.

The remaining coupling $\widetilde{S}_B$, which multiplies the flavour-singlet structure that couples to the topological charge density $\omega$, is not fixed by the quark axial charges: it parametrises the part of the singlet axial current that is not determined by the anomalous Ward identity. It can therefore be extracted only from observables sensitive to the matrix element of $\omega$ between nucleon states, 
see Appendix~\ref{app:stilde_extraction}.

\subsection{The light ALP limit}
\label{sec:light-ALP}

The general couplings in Eq.~\eqref{eq:baryoncouplings} simplify in the light ALP limit, namely $m_a\ll m_\pi$. Recalling the large-$m_0$ expression for $\mathbb M_a^{-1}(\bm{\mathcal C})$ in Eq.~\eqref{eq:Ma_light_large_m0}, and neglecting corrections suppressed by $m_{u,d}/m_s$, the flavour-conserving axial couplings are
\begin{equation}
\begin{split}
c^A_{pp} &=c_{GP} \frac{m_u {\Delta}d + m_d {\Delta}u}{m_u + m_d} + c_{uu}^{AP} {\Delta}u + 
c_{dd}^{AP} {\Delta}d + c_{ss}^{AP} {\Delta}s,\\
c^A_{nn} &=c_{GP} \frac{m_u {\Delta}u + m_d {\Delta}d}{m_u + m_d} + c_{uu}^{AP} {\Delta}d + 
c_{dd}^{AP} {\Delta}u + c_{ss}^{AP} {\Delta}s,\\
c^A_{\Lambda\Lambda} &=(c_{GP}+c_{uu}^{AP}+c_{dd}^{AP})\frac{{\Delta}u+4{\Delta}d+{\Delta}s}{6}
+c_{ss}^{AP}\frac{2{\Delta}u-{\Delta}d+2{\Delta}s}{3},\\
c^A_{\Sigma^+\Sigma^+} &=c_{GP}\frac{m_u{\Delta}s+m_d{\Delta}u}{m_u+m_d}
+c_{uu}^{AP}{\Delta}u+c_{dd}^{AP}{\Delta}s+c_{ss}^{AP}{\Delta}d,\\
c^A_{\Sigma^0\Sigma^0} &=(c_{GP}+c_{uu}^{AP}+c_{dd}^{AP})\frac{{\Delta}u+{\Delta}s}{2}
+c_{ss}^{AP}{\Delta}d,\\
c^A_{\Sigma^-\Sigma^-} &=c_{GP}\frac{m_u{\Delta}u+m_d{\Delta}s}{m_u+m_d}
+c_{uu}^{AP}{\Delta}s+c_{dd}^{AP}{\Delta}u+c_{ss}^{AP}{\Delta}d,\\
c^A_{\Xi^0\Xi^0} &=c_{GP}\frac{m_u{\Delta}s+m_d{\Delta}d}{m_u+m_d}
+c_{uu}^{AP}{\Delta}d+c_{dd}^{AP}{\Delta}s+c_{ss}^{AP}{\Delta}u,\\
c^A_{\Xi^-\Xi^-} &=c_{GP}\frac{m_u{\Delta}d+m_d{\Delta}s}{m_u+m_d}
+c_{uu}^{AP}{\Delta}s+c_{dd}^{AP}{\Delta}d+c_{ss}^{AP}{\Delta}u,\\
c^A_{\Sigma^0\Lambda} & =c^A_{\Lambda\Sigma^0} =
    \frac{{\Delta}u-2{\Delta}d+{\Delta}s}{2\sqrt{3}}
    \left(c_{GP}\frac{m_d-m_u}{m_u+m_d}+c_{uu}^{AP}-c_{dd}^{AP}\right),
\end{split}
\label{eq:effcoupl_fc}
\end{equation}
The dependence on the $\widetilde{S}_B$ coefficient vanishes in the light ALP limit, as $\partial_\mu X_\mathrm{TC} \sim \mathcal{O}(m_a^2/m_\pi^2)$ by combining Eq.~\eqref{eq:baryon_ALP_building_blocks} and~\eqref{eq:minv_low_ma}.
The proton and neutron couplings reproduce the results of Refs.~\cite{GrillidiCortona:2015jxo,DiLuzio:2020wdo,Bauer:2021mvw}.

In the same limit, the flavour-changing couplings are
\begin{equation}\begin{aligned}
c^A_{p\Sigma^+} &= c_{ds}^{AP}({\Delta}s - {\Delta}d),\qquad& c^V_{p \Sigma^+} &= -c_{ds}^{VS}\\
 c^A_{n\Sigma^0} &= \frac{c_{ds}^{AP}}{\sqrt{2}}({\Delta}d-{\Delta}s),\qquad& c^V_{n \Sigma^0} 
 &= \frac{c_{ds}^{VS}}{\sqrt{2}},\\
 c^A_{n\Lambda} &= \frac{c_{ds}^{AP}}{\sqrt{6}}(-2{\Delta}u+{\Delta}d + {\Delta}s),\qquad& c^V_{n \Lambda} 
 &= -\sqrt{\frac{3}{2}}c_{ds}^{VS},\\
 c_{\Sigma^0 \Xi^0}^A &= \frac{c_{ds}^{AP}}{\sqrt{2}}({\Delta}d-{\Delta}u),\qquad& c^V_{\Sigma^0 \Xi^0} 
 &= -\frac{c_{ds}^{VS}}{\sqrt{2}}, \\
c_{\Sigma^- \Xi^-}^A &= c_{ds}^{AP}({\Delta}u-{\Delta}d),&\qquad c^V_{\Sigma^- \Xi^-} &= c_{ds}^{VS},\\
c_{\Lambda \Xi^0}^A &= \frac{c_{ds}^{AP}}{\sqrt{6}}({\Delta}u+{\Delta}d-2{\Delta}s),\qquad&c^V_{\Lambda \Xi^0} 
 &=\sqrt{\frac{3}{2}}c_{ds}^{VS}.
\label{eq:effcoupl_fv}\end{aligned}\end{equation}
Unlike the flavour-conserving couplings, these expressions are independent of both singlet coefficients for all values of $m_a$. They satisfy
$c_{B_2B_1}^{A,V} =\left(c_{B_1B_2}^{A,V}\right)^*$,
as required by hermiticity.

\subsection{Continuation to the GeV scale}
\label{sec:baryon_GeV_continuation}

The general couplings in Eq.~\eqref{eq:baryoncouplings} encode their dependence on the ALP mass through
$\mathbb M_a^{-1}(\bm{\mathcal C})$. Within the domain of validity of the chiral theory, this expression 
directly describes the variation of the couplings with $m_a$. The poles occurring when the
ALP mass approaches a neutral pseudoscalar mass signal the breakdown of the perturbative treatment of 
ALP--meson mixing. The corresponding expressions should therefore not be used in the nearly degenerate
regions.

The phenomenological study of the ALP decays $a\to B_1 \overline B_2$ requires $m_a\geq m_{B_1}+m_{B_2}$ 
and consequently lies beyond the regime in which the chiral description alone is reliable. To connect the
chiral result to this region, we introduce a phenomenological continuation constrained by the asymptotic
scaling of baryonic form factors according to hard-scattering counting~\cite{Brodsky:1974vy,Lepage:1980fj,Brodsky:1980sx}. This prescription does not
extend the validity of chiral perturbation theory, but provides a model for the effective ALP--baryon
couplings at masses of a few GeV.

Consider the decay of an ALP with momentum $q$ into on-shell baryons $B_1(p_1)$ and 
$\bar{B}_2(p_2)$. We define, for convenience, the quark bilinears and the topological gluon operators, as
\begin{equation}
V_{ij}^\mu\equiv\bar q_i\gamma^\mu q_j, \qquad
A_{ij}^\mu\equiv\bar q_i\gamma^\mu\gamma_5 q_j, \qquad
P_{ij}\equiv\bar q_i i\gamma_5 q_j, \qquad
S_{ij}\equiv\bar q_i q_j.
\label{eq:general_baryon_operator_definitions}
\end{equation}
Their matrix elements admit the following decomposition in terms of Lorentz invariant form factors:
\begin{equation}
\begin{aligned}
\langle B_1(p_1)\bar B_2(p_2)|V_{ij}^\mu|0\rangle
&=\bar u_1 \left[ \gamma^\mu F_{1}^{ij}(q^2)
+\frac{i\sigma^{\mu\nu}q_\nu}{m_{B_1}+m_{B_2}}F_{2}^{ij}(q^2)
+\frac{q^\mu}{m_{B_1}+m_{B_2}}F_{3}^{ij}(q^2)\right]v_2,\\
\langle B_1(p_1)\bar B_2(p_2)|A_{ij}^\mu|0\rangle
&= \bar u_1 \left[ \gamma^\mu G_{A}^{ij}(q^2)
+\frac{i\sigma^{\mu\nu}q_\nu}{m_{B_1}+m_{B_2}}G_{T}^{ij}(q^2)
+\frac{q^\mu}{m_{B_1}+m_{B_2}}G_{P}^{ij}(q^2)
\right]\gamma_5v_2,\\
\langle B_1(p_1)\bar B_2(p_2)|P_{ij}|0\rangle
&=G_{5}^{ij}(q^2)\,\bar u_1i\gamma_5v_2,\\
\langle B_1(p_1)\bar B_2(p_2)|S_{ij}|0\rangle
&=F_{S}^{ij}(q^2)\,\bar u_1v_2,\\
\langle B_1(p_1)\bar B_2(p_2)|
\omega|0\rangle
&=\frac{1}{2}(m_{B_1}+m_{B_2}) G_{\widetilde G}(q^2)\,
\bar u_1i\gamma_5v_2,
\end{aligned}
\label{eq:general_baryon_pair_form_factors}
\end{equation}
with $u_1\equiv u_{B_1}(p_1)$ and $v_2\equiv v_{B_2}(p_2)$ the spinors associated with the baryonic final state, and $\omega$ is the topological charge density defined in Eq.~\eqref{eq:topo_density}.
All form factors in this 
convention are dimensionless and may be complex in the timelike region. 
The longitudinal matrix elements are obtained by contracting the vector and axial-vector matrix elements with the final total momentum $q=p_1+p_2$,
\begin{equation}
\begin{aligned}
q_\mu\langle B_1\bar B_2|V_{ij}^\mu|0\rangle
&=
\left[
(m_{B_1}-m_{B_2})F_{1}^{ij}(q^2)
+\frac{q^2}{m_{B_1}+m_{B_2}}F_{3}^{ij}(q^2)
\right]\bar u_1v_2,
\\
q_\mu\langle B_1\bar B_2|A_{ij}^\mu|0\rangle
&=
(m_{B_1}+m_{B_2}) D^{ij}(q^2)\,\bar u_1\gamma_5v_2,
\end{aligned}
\label{eq:general_baryon_longitudinal_form_factors}
\end{equation}
with 
\begin{equation}
D^{ij}(q^2)
\equiv
G_{A}^{ij}(q^2)
+\frac{q^2}{(m_{B_1}+m_{B_2})^2}G_{P}^{ij}(q^2).
\end{equation}
Thus neither the Pauli form factor $F_{2}^{ij}$ nor the
induced pseudotensor form factor $G_{T}^{ij}$ contributes
to the longitudinal amplitudes.

With the anomaly sign adopted in this work, the vector and axial vector quark current Ward identities read
\begin{equation}
\begin{aligned}
\partial_\mu V_{ij}^\mu
&=i(m_{q_i}-m_{q_j})S_{ij},
\\
\partial_\mu A_{ij}^\mu
&=(m_{q_i}+m_{q_j})P_{ij}
-2\omega\,\delta_{ij},
\end{aligned}
\label{eq:general_quark_ward_identities}
\end{equation}
implying the following relation between the baryonic form factors:
\begin{equation}
\begin{aligned}
(m_{q_i}-m_{q_j})\, F_{S}^{ij}(q^2) = {}& 
(m_{B_1}-m_{B_2})\, F_{1}^{ij}(q^2)
+\frac{q^2}{m_{B_1}+m_{B_2}}\,F_{3}^{ij}(q^2), \\
(m_{q_i}+m_{q_j})\, G_{5}^{ij}(q^2) = {}&
(m_{B_1}+m_{B_2}) \left(D^{ij}(q^2)+\delta_{ij}G_{\widetilde G}(q^2)\right).
\end{aligned}
\label{eq:general_baryon_form_factor_ward_identities}
\end{equation}
The Ward identities constrain the longitudinal form factors. For a
flavour-changing vector current, $m_{q_i}\neq m_{q_j}$, the vector
identity expresses the scalar form factor in terms of the longitudinal
vector combination. For an identical $B\bar B$ final state, flavour
conservation allows only flavour-diagonal vector currents, whose longitudinal matrix elements vanish by current conservation.
Since $m_{B_1}=m_{B_2}$, this condition gives
$F_3(q^2)=0$ for $q^2\neq0$, and the derivative vector coupling to baryons therefore
does not contribute to the on-shell $a\to B\bar B$ amplitude.

The axial-vector identity expresses the corresponding longitudinal
combination in terms of the pseudoscalar and topological form factors.
Its nonsinglet projection is simple because the
gluonic anomaly cancels. For coefficients $t_{ij}$ satisfying $\sum t_{ii} = 0$, we define $G_X^{\rm NS}=\sum t_{ij}G_X^{ij}$
for $X=A,T,P$ and
$D_{\rm NS}=G_A^{\rm NS}+q^2G_P^{\rm NS}/(m_{B_1}+m_{B_2})^2$. In the
chiral limit, the Ward identity reduces to current conservation and implies
\begin{equation}
D_{\mathrm{NS}}(q^2)=0,
\qquad
G_P^{\mathrm{NS}}(q^2)
=
-\frac{(m_{B_1}+m_{B_2})^2}{q^2}
G_A^{\mathrm{NS}}(q^2).
\label{eq:nonsinglet_axial_conservation}
\end{equation}
Combined with hard scattering counting, this gives
\begin{equation}
\begin{aligned}
G_A^{\mathrm{NS}}(q^2)
&\propto
\frac{\alpha_s^2(-q^2)}{(-q^2)^2},
\\
G_P^{\mathrm{NS}}(q^2)
&\propto
\frac{(m_{B_1}+m_{B_2})^2\alpha_s^2(-q^2)}
{(-q^2)^3},
\end{aligned}
\label{eq:brodskylepage}
\end{equation}
in the $q^2\to-\infty$ limit, up to logarithmic corrections and overall dimensionful normalisation factors
\cite{Brodsky:1974vy,Lepage:1980fj,Brodsky:1980sx}.
The two contributions to $D_{\mathrm{NS}}(q^2)$ consequently have
the same asymptotic power and cancel, as required by current
conservation.

At finite quark masses, or for a singlet axial current, the
longitudinal amplitude does not vanish and is related by
Eq.~\eqref{eq:general_baryon_form_factor_ward_identities} to the
pseudoscalar and topological form factors. The hard
contributions considered here behave as
\begin{equation}
G_5^{ij}(q^2)
\propto
\frac{\alpha_s^2(-q^2)}{(-q^2)^2},
\qquad
G_{\widetilde G}(q^2)
\propto
\frac{\alpha_s^3(-q^2)}{(-q^2)^2},
\label{eq:axial_longitudinal_asymptotic}
\end{equation}
in the $q^2\to-\infty$ limit. 
The scaling of the pseudoscalar form factor follows from hard scattering counting
\cite{RuizArriola:2025omi,Chua:2002yd}. For the topological form
factor, we adapt the three valence quark calculation of
Ref.~\cite{Tong:2022zax}, replacing the two gluon scalar insertion
by
\begin{equation}
V_P^{\mu\nu}(l,r)
=
\epsilon^{\mu\nu\rho\sigma}l_\rho r_\sigma.
\end{equation}
Here $l$ and $r$ are the gluon momenta.
The collinear contribution vanishes, while the term linear in the
transverse parton momentum produces a nonvanishing
twist-3--twist-4 overlap. Four quark--gluon vertices provide a factor
$g_s^4\propto\alpha_s^2$, and the explicit coupling in
$\mathcal O_{\widetilde G}$ provides an additional power of
$\alpha_s$. This calculation motivates the hard scaling in
Eq.~\eqref{eq:axial_longitudinal_asymptotic}, but still does not determine
its normalisation, endpoint logarithms, or the scale at which this
contribution becomes dominant.

In the flavour-changing vector sector, hard-scattering counting gives
\begin{equation}
F_1^{ij}(q^2)
\propto
\frac{\alpha_s^2(-q^2)}{(-q^2)^2},
\qquad
F_3^{ij}(q^2)
\propto
\frac{\alpha_s^2(-q^2)}{(-q^2)^3},
\label{eq:vector_form_factor_asymptotic}
\end{equation}
in the $q^2\to-\infty$ limit. Here too, the dimensionful proportionality factors required by the
different form factor dimensions are implicit
\cite{Belitsky:2002kj}. Consequently, the two terms in the
longitudinal combination
\begin{equation}
(m_{B_1}-m_{B_2})F_1^{ij}(q^2)
+
\frac{q^2}{m_{B_1}+m_{B_2}}F_3^{ij}(q^2)
\end{equation}
have the same
$\alpha_s^2(-q^2)/(-q^2)^2$ behaviour. For
$m_{q_i}\neq m_{q_j}$, the vector Ward identity then implies the
same asymptotic power for the corresponding scalar form factor
$F_S^{ij}(q^2)$.

We now match the QCD form factors onto the effective ALP--baryon
couplings. The coefficients in Eq.~\eqref{eq:baryoncoupl_full} are interpreted
as momentum-dependent on-shell vertex functions evaluated at $q^2=m_a^2$.
For the decay
$a\to B_1\overline B_2$, the corresponding matrix element is
\begin{eqnarray}
\langle B_1\overline B_2|\mathcal L_{aB}|0\rangle &=&
-i\frac{a}{f_a}q_\mu\,
\overline u_1\gamma^\mu
\left[
c_{B_1B_2}^V(q^2)
+c_{B_1B_2}^A(q^2)\gamma_5
\right]v_2 \nonumber 
\\
& & \hspace{-2.5cm} =
-\frac{a}{f_a}
\bigg[
i(m_{B_1}-m_{B_2})c_{B_1B_2}^V(q^2)
\,\overline u_1v_2
+(m_{B_1}+m_{B_2})c_{B_1B_2}^A(q^2)
\,\overline u_1i\gamma_5v_2
\bigg].
\label{eq:general_onshell_baryon_matching}
\end{eqnarray}
Using the partonic ALP Lagrangian in
Eq.~\eqref{eq:lagr_partons}, together with the form factor
decompositions and Ward identities given above, and matching the
independent scalar and pseudoscalar spinor structures, we obtain
\begin{equation}
\begin{aligned}
(m_{B_1}-m_{B_2})c_{B_1B_2}^V(q^2)
&=
\sum_{i,j}(m_{q_i}-m_{q_j})
c_{ij}^{VS}F_S^{ij}(q^2),
\\
c_{B_1B_2}^A(q^2)
&=
\sum_{i,j}c_{ij}^{AP}
\frac{m_{q_i}+m_{q_j}}{m_{B_1}+m_{B_2}}
G_5^{ij}(q^2)
-c_{GA}G_{\widetilde G}(q^2).
\end{aligned}
\label{eq:general_baryon_coupling_form_factors}
\end{equation}
The dependence on the partonic ALP coefficients is therefore entirely
expressed through the field redefinition invariant combinations
introduced in Eq.~\eqref{eq:invariants}. For
$m_{B_1}\neq m_{B_2}$, the first relation determines
$c_{B_1B_2}^V(q^2)$. For equal baryon masses, the derivative vector
interaction vanishes on-shell and its coefficient cannot be determined
from this amplitude.

Equation~\eqref{eq:general_baryon_coupling_form_factors}
identifies the invariant QCD form factors governing the on-shell
ALP--baryon couplings, but does not determine their nonperturbative
normalisation in the low-energy region. We therefore use the chiral
couplings in Eq.~\eqref{eq:baryoncouplings} as the low-energy input
for the phenomenological continuation and dress their invariant source
contributions according to the asymptotic form factor scaling discussed
above.

To define the dressing factors, we first identify the powers of $q^2$
already present in the large $q^2$ expansion of the chiral
building blocks. We decompose the components linear in the ALP field 
in Eq.~\eqref{eq:baryon_ALP_building_blocks} into contributions
proportional to $\bm{c_q^{AP}}$ and $c_{GA}$, denoted by the
superscripts $AP$ and $GA$, respectively. Their asymptotic
limits, obtained using Eq.~\eqref{eq:Ma_heavy_limit}, are
\begin{equation}
\begin{aligned}
\left.\bm{u_\mu^{AP}}\right|_a
&\simeq
-2\frac{\partial_\mu a}{f_a}
\frac{B_0\{\bm{m_q},\bm{c_q^{AP}}\}}{q^2},
&
\left.\bm{u_\mu^{GA}}\right|_a
&\simeq
-2\frac{\partial_\mu a}{f_a}
\frac{m_0^2}{N_fq^2}c_{GA}\bm 1,
\\
\left.\partial_\mu X_{TC}^{AP}\right|_a
&\simeq
-2\frac{\partial_\mu a}{f_a}
\frac{B_0
\left\langle\{\bm{m_q},\bm{c_q^{AP}}\}\right\rangle}{q^2},
&
\left.\partial_\mu X_{TC}^{GA}\right|_a
&\simeq
-2\frac{\partial_\mu a}{f_a}
\frac{m_0^2+q^2}{q^2}c_{GA}.
\end{aligned}
\label{eq:heavy_ALP_building_blocks}
\end{equation}
Thus, the contributions proportional to $\bm{c_q^{AP}}$, together
with the $c_{GA}$ contribution to $\bm{u_\mu}$, already scale as
$1/q^2$. By contrast, the $c_{GA}$ contribution to
$\partial_\mu X_{TC}$ approaches a constant. The vector contribution
in Eq.~\eqref{eq:baryon_ALP_building_blocks} is also independent of
$q^2$. These different behaviours determine the additional
suppression that must be assigned to each contribution in order to
reproduce the asymptotic form factor scaling.

We implement the continuation in the timelike region by introducing a
matching scale $\mu_\ast$ and defining
\begin{equation}
\mathcal R_{n,k}(q^2)=
\begin{cases}
1,
& q^2\leq\mu_\ast^2,
\\[1mm]
\displaystyle
\left[
\frac{\alpha_s(q^2)}
{\alpha_s(\mu_\ast^2)}
\right]^n
\left(\frac{\mu_\ast^2}{q^2}\right)^k,
& q^2>\mu_\ast^2.
\end{cases}
\label{eq:baryon_continuation_profile}
\end{equation}
The indices $n$ and $k$ account, respectively, for the powers of
$\alpha_s$ and for the additional power suppression required beyond
the behaviour already contained in the chiral building blocks.
According to the asymptotic scaling derived above, we prescribe
\begin{equation}
\begin{aligned}
\left.\bm{u_\mu}\right|_a
&\longrightarrow
\mathcal R_{2,1}(q^2)
\left.\bm{u_\mu^{AP}}\right|_a
+
\mathcal R_{3,1}(q^2)
\left.\bm{u_\mu^{GA}}\right|_a,
\\
\left.\partial_\mu X_{TC}\right|_a
&\longrightarrow
\mathcal R_{2,1}(q^2)
\left.\partial_\mu X_{TC}^{AP}\right|_a
+
\mathcal R_{3,2}(q^2)
\left.\partial_\mu X_{TC}^{GA}\right|_a,
\\
\left.\bm{\nabla_\mu}\bm{\mathcal B}\right|_a
&\longrightarrow
\mathcal R_{2,2}(q^2)
\left.\bm{\nabla_\mu}\bm{\mathcal B}\right|_a.
\end{aligned}
\label{eq:baryon_source_dressing}
\end{equation}
The factors act on the exact invariant source contributions, rather
than on their large $q^2$ approximations or on individual chiral
coefficients.  In particular, the different factors
multiplying the two $c_{GA}$ contributions compensate for their
different initial powers in Eq.~\eqref{eq:heavy_ALP_building_blocks}
and produce the same
$\alpha_s^3(q^2)/(q^2)^2$ asymptotic behaviour.

Neglecting the light quark masses relative to $m_s$, the diagonal
axial couplings at large ALP mass, with $q^2=m_a^2$, take the compact
form
\begin{equation}
\begin{aligned}
c_{BB}^A\simeq{}&
c_{ss}^{AP}
\frac{B_0m_s}{m_a^2}
x_{BB}\,\mathcal R_{2,1}(m_a^2)-
c_{GA}
\sum_{q=u,d,s} \!\!{\Delta} q
\frac{m_0^2}{N_fm_a^2}
\mathcal R_{3,1}(m_a^2)
-
c_{GA}\widetilde S_B\,
\mathcal R_{3,2}(m_a^2),
\end{aligned}
\label{eq:baryon_high_energy_dressed}
\end{equation}
where the flavour coefficients are
\begin{equation}
\begin{aligned}
x_{pp}=x_{nn}
&=\Delta s + \widetilde{S}_B,\qquad
&
x_{\Lambda\Lambda}
&=
\frac{2\Delta u-\Delta d+2\Delta s}{3}+ \widetilde{S}_B,
\\
x_{\Sigma^+\Sigma^+}
=x_{\Sigma^0\Sigma^0}
=x_{\Sigma^-\Sigma^-}
&=\Delta d+ \widetilde{S}_B,
&
x_{\Xi^0\Xi^0}
&=x_{\Xi^-\Xi^-}
=\Delta u+ \widetilde{S}_B.
\end{aligned}
\label{eq:baryon_high_energy_flavour}
\end{equation}

Equations~\eqref{eq:baryon_continuation_profile}--\eqref{eq:baryon_high_energy_flavour}
define our phenomenological continuation of the chiral ALP--baryon
couplings. The prescription is continuous at the matching scale,
although its derivative need not be. The asymptotic form factor
scaling derived in the spacelike region motivates the timelike
envelopes, but does not determine the matching scale, the relative
normalisations, or the generally complex phases of the timelike form
factors. Possible effects from distribution amplitude evolution and
endpoint logarithms are also not resolved by this parametrisation.

The anomalous Ward identity constrains the complete matrix elements,
but does not uniquely determine the separate dressing factors assigned
to the different chiral source contributions. The prescription should
therefore be regarded as a model constrained by the asymptotic QCD
scaling, rather than as an extension of the validity of chiral
perturbation theory. Predictions for ALP masses of a few GeV consequently
retain the theoretical uncertainties associated with this continuation.
In the following subsection, we use this prescription to study the
baryon couplings across the ALP mass range.

\subsection{Baryon couplings across the ALP mass range}
\label{sec:baryon_couplings_mass}

We now use the results of the previous subsections to study the baryon
couplings as functions of the ALP mass, from the light ALP limit to the
region of a few GeV. 
For the numerical analysis, we use the proton quark axial charges
determined in Ref.~\cite{Alexandrou:2024ozj}, renormalised in the
$\overline{\mathrm{MS}}$ scheme at $\mu=2\,\mathrm{GeV}$,
\begin{equation}
\Delta u=0.832(28),\qquad
\Delta d=-0.417(22),\qquad
\Delta s=-0.037(18).
\label{eq:axial_charges_lattice}
\end{equation}
Using the matching relations in Eq.~\eqref{eq:axial_matching} and
neglecting correlations among the quoted single-flavour uncertainties,
these values give
\begin{equation}
D_B=0.815(28),\qquad
F_B=0.434(17),\qquad
S_B=-0.417(22).
\label{eq:baryon_LECs}
\end{equation}

A procedure to estimate $\widetilde{S}_B$, at leading order in $\mathcal{O}(\alpha_s)$ and within the approximation of pointlike baryon, is described in Appendix~\ref{app:stilde_extraction}. We obtained
\begin{equation}
    \widetilde S_B = 0.025\pm0.031\quad(\text{na\"ive}),\qquad \widetilde S_B = 0.025\pm0.062\quad(\text{conservative}).\label{eq:stilde_value}
\end{equation}
We have adopted two different prescriptions, na\"ive or conservative, to estimate its phenomenological range.

\begin{figure}
\centering
\includegraphics[scale=0.92]{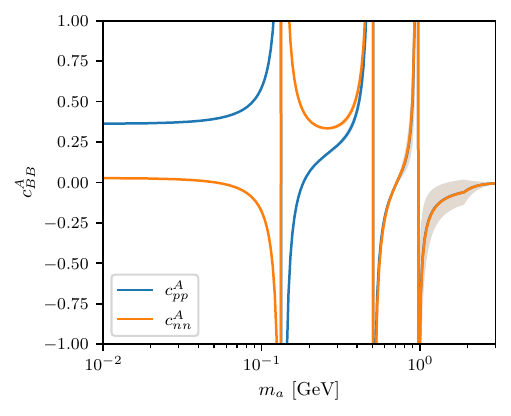}
\includegraphics[scale=0.92]{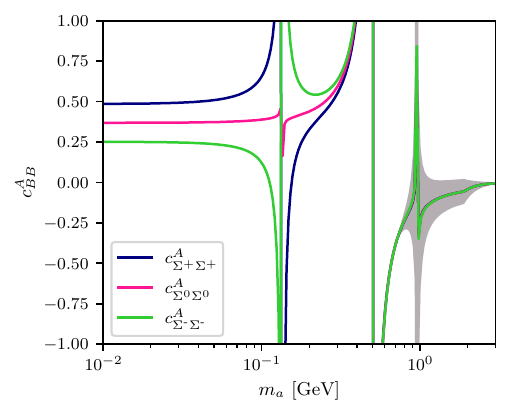}\\
\includegraphics[scale=0.92]{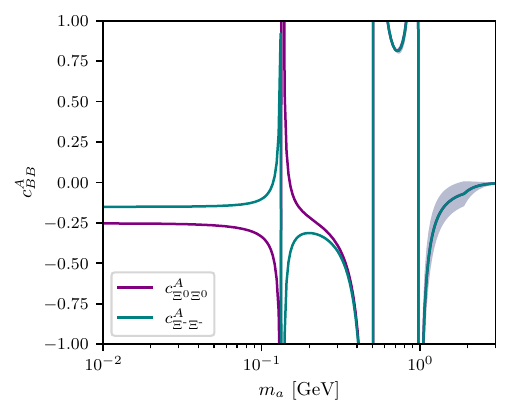}
\includegraphics[scale=0.92]{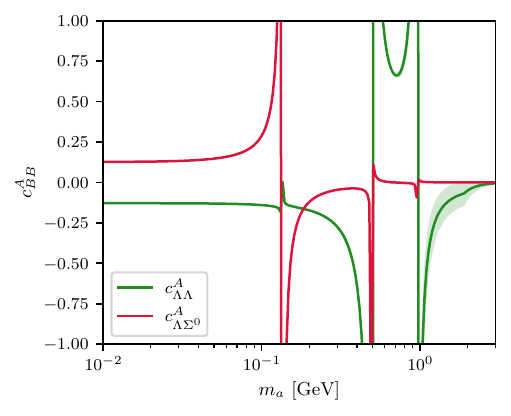}
\caption{Effective flavour-conserving axial couplings in the
gluophilic benchmark ($c_{GA} = c_{GP}=1$, $\bm{c_q^{AP}} = \bm{c_q^{VS}}= 0$) with $\mu_\star = 1.9\,\mathrm{GeV}$. The solid curves correspond to $\widetilde S_B=0.025$. 
The shaded regions are obtained by varying $\widetilde S_B$ within the conservative prescription $\widetilde{S}_B = 0.025\pm 0.062$.
Top left: nucleons; top right: $\Sigma$ baryons; bottom left:
$\Xi$ baryons; bottom right: the $\Lambda$ baryon and the
$\Sigma^0$--$\Lambda$ transition.}
\label{fig:effcoupl_fc}
\end{figure}

We first consider flavour-conserving quark interactions. In this case,
the coefficients $c_{BB}^V$ do not contribute to the on-shell amplitudes and 
the relevant couplings are therefore the axial couplings $c_{BB}^A$ and 
$c_{\Sigma^0\Lambda}^A = c_{\Lambda\Sigma^0}^A$.
To illustrate their dependence on the ALP mass, we consider the
gluophilic benchmark defined in terms of the invariant coefficients by
\begin{equation}
c_{GA}=c_{GP}=1,\qquad
\bm{c_q^{AP}}=\bm{c_q^{VS}}=0.
\label{eq:gluophilic_benchmark}
\end{equation}

The pole structure visible in Fig.~\ref{fig:effcoupl_fc} reflects the
neutral pseudoscalar spectrum contained in
$\mathbb M_a^{-1}(\bm{\mathcal C})$. As discussed above, the perturbative treatment is not reliable when the ALP is nearly degenerate with one of the pseudoscalar mesons. The coloured bands in the figures are obtained by varying $\widetilde{S}_B$ within a range given by three times its estimated uncertainty of Eq.~\eqref{eq:stilde_value}. The second singlet operator is most impactful when $m_a \gtrsim m_{\eta^\prime}$...

In the gluophilic benchmark, the ALP source is isoscalar. Differences
among members of the same isospin multiplet arise from the
isospin-violating part of
$\mathbb M_a^{-1}(\bm{\mathcal C})$, which is dominated in this mass
range by the physical $\pi^0$ pole. For ALP masses sufficiently above
$m_{\pi^0}$, the pion-pole contribution is suppressed and the residual
isospin breaking is numerically negligible. The remaining isoscalar 
contributions then lead to the approximate relations
\begin{equation}
c_{pp}^A=c_{nn}^A,\qquad
c_{\Sigma^0\Sigma^0}^A
=c_{\Sigma^+\Sigma^+}^A
=c_{\Sigma^-\Sigma^-}^A,\qquad
c_{\Xi^0\Xi^0}^A=c_{\Xi^-\Xi^-}^A,
\end{equation}
as can be seen in Fig.~\ref{fig:effcoupl_fc}. By contrast, $c_{\Sigma^0\Lambda}^A$ 
depends only on the isovector component and is suppressed in this regime.

\begin{figure}
\centering
\includegraphics[scale=0.92]{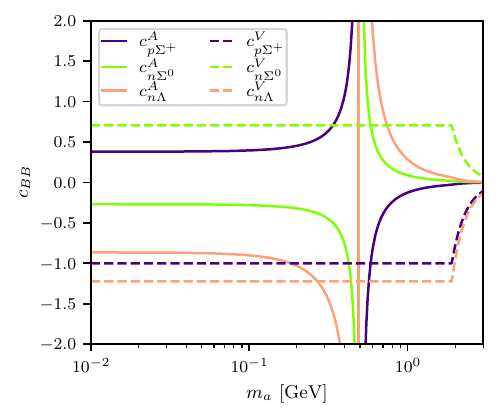}
\includegraphics[scale=0.92]{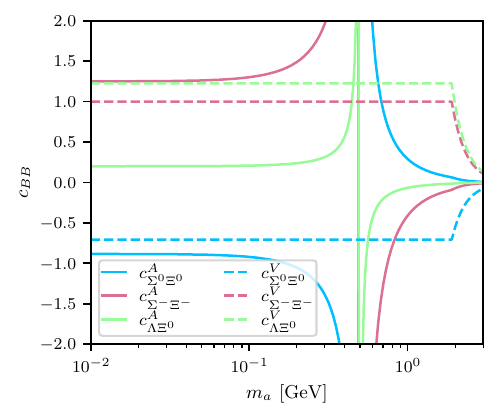}
\caption{Effective flavour-changing baryon couplings with $\mu_\star = 1.9\,\mathrm{GeV}$. The solid curves correspond
to $c_{ds}^{AP}=1$, while the dashed curves correspond to
$c_{ds}^{VS}=1$. Left: transitions involving nucleons. Right:
transitions involving only hyperons. The axial couplings depend on the
nonsinglet coefficients $D_B$ and $F_B$, while the vector couplings are
fixed by the flavour structure of $\bm{c_q^{VS}}$. All the couplings
shown are independent of $S_B$ and $\widetilde S_B$.}
\label{fig:effcoupl_fv_section}
\end{figure}

We next consider flavour-changing interactions in the $s\to d$ sector.
The two benchmark scenarios are defined by $\left(c_{ds}^{AP},c_{ds}^{VS}\right)=(1,0)$ 
or $\left(c_{ds}^{AP},c_{ds}^{VS}\right)=(0,1)$,
with the conjugate coefficients fixed by hermiticity and all remaining
invariant coefficients set to zero. The first scenario generates the
axial couplings, whose dependence on $m_a$ is governed by
$\mathbb M_a^{-1}(\bm{\mathcal C})$. The second generates the vector
couplings, which are independent of $m_a$ within the chiral
description and acquire a mass dependence only through the form factor
introduced in Eq.~\eqref{eq:baryon_source_dressing}.

The pole in the axial couplings at $m_a=m_{K^0}$ reflects the pseudoscalar 
spectrum entering $\mathbb M_a^{-1}(\bm{\mathcal C})$ and marks the region 
in which the perturbative treatment is not applicable. Away from this region, 
the large-mass behaviour of the axial couplings is
\begin{equation}
\left.c_{B_1B_2}^A\right|_{m_a\gg m_{K^0}} \simeq
-\frac{m_{K^0}^2}{m_a^2} \mathcal R_{2,1}(m_a^2) \left.c_{B_1B_2}^A\right|_{m_a=0}.
\end{equation}
The vector couplings are independent of $m_a$ in the chiral description and 
acquire their mass dependence through $\mathcal R_{2,2}(m_a^2)$ in
Eq.~\eqref{eq:baryon_source_dressing}:
\begin{equation}
c_{B_1B_2}^V(m_a^2)
=
\mathcal R_{2,2}(m_a^2)c_{B_1B_2}^V(0).
\end{equation}

Ref.~\cite{DiLuzio:2023cuk} derived the derivative ALP couplings
    to the baryon octet, including the flavour-changing couplings in the
$s\to d$ sector, within an $SU(3)$ chiral framework in which the
$\eta'$ contribution and the independent operator proportional to
$\widetilde S_B$ are omitted. In the common regime
$m_a\ll m_{\eta'}$ and for $\widetilde S_B=0$, our expressions agree
with their results, after accounting for the different normalisation
conventions. Ref.~\cite{Blinov:2021say} instead employed a $U(3)$
description including the $\eta'$ to determine the ALP--proton
coupling, but likewise did not include the independent interaction
proportional to $\widetilde S_B$. Our construction treats the complete
baryon octet with the full neutral pseudoscalar sector, includes this
additional singlet operator, and provides a phenomenological
continuation of the couplings towards the few-GeV region.

The flavour-changing couplings provide the input for the light ALP
baryon decays studied in Sec.~\ref{sec:lightpheno}, whereas the
diagonal axial couplings determine the baryon pair channels discussed
in Sec.~\ref{sec:baryonpairs}.

\section{Light ALP probes in flavour-changing baryon decays}
\label{sec:lightpheno}

The flavour-changing couplings derived in Eq.~\eqref{eq:effcoupl_fv} generate
the two-body decays $B_1\to B_2a$ whenever $m_a<m_{B_1}-m_{B_2}$. The small 
mass splittings within the baryon octet restrict these processes to light
ALPs, with the largest kinematically accessible masses of order $250\,\mathrm{MeV}$.
These decays probe both the vector and axial $s\to d$ interactions and are 
therefore complementary to pseudoscalar meson decays, which are insensitive 
to the corresponding axial current.

A long-lived ALP, or an ALP that decays into invisible states, produces
the signature $B_1\to B_2+\mathrm{inv}$. If the ALP instead decays
promptly into SM particles, it can be searched for as a resonance in
the visible final state. We consider below the invisible searches
performed by BESIII and the prompt dimuon channel studied by LHCb.

For the interaction defined in Eq.~\eqref{eq:baryoncoupl_full}, with
the couplings evaluated at $q^2=m_a^2$, the decay width is 
\begin{equation}
\begin{aligned}
\Gamma(B_1\to B_2a)
={}&
\frac{m_a^4\sqrt{z_+z_-}}
{16\pi m_{B_1}^3f_a^2}
\bigg[
|c_{B_1B_2}^V|^2
(m_{B_1}-m_{B_2})^2z_+
+
|c_{B_1B_2}^A|^2
(m_{B_1}+m_{B_2})^2z_-
\bigg],
\end{aligned}
\label{eq:baryon_ALP_decay_width}
\end{equation}
where $z_\pm =(m_{B_1}\pm m_{B_2})^2/m_a^2-1$. For $m_a\ll (m_{B_1}-m_{B_2})$, 
Eq.~\eqref{eq:baryon_ALP_decay_width} reduces to
\begin{equation}
\Gamma(B_1\to B_2a) \simeq
\frac{m_{B_1}^3}{16\pi f_a^2} \left( |c_{B_1B_2}^V|^2+|c_{B_1B_2}^A|^2\right)
\left(1-\frac{m_{B_2}^2}{m_{B_1}^2}\right)^3.
\label{eq:baryon_ALP_decay_width_light}
\end{equation}

\subsection{Invisible ALP searches}

BESIII has searched for the two-body decays
$\Sigma^+\to p+\mathrm{inv}$~\cite{BESIII:2023utd} and
$\Xi^0\to\Lambda+\mathrm{inv}$~\cite{BESIII:2026ztz}. Both analyses
assume a massless invisible particle escaping the detector, so we apply their
results in the limit $m_a\simeq0$.

\begin{figure}
\centering
\includegraphics[width=0.7\textwidth]{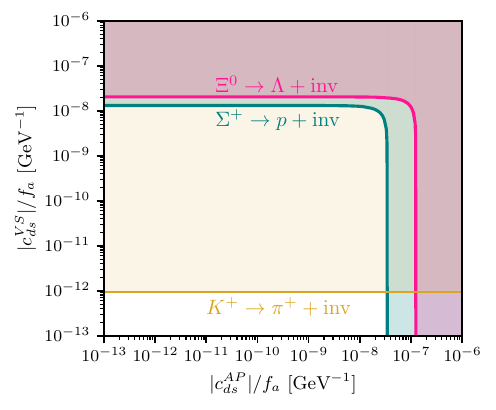}
\caption{Constraints on the invariant flavour-changing coefficients
$c_{ds}^{AP}$ and $c_{ds}^{VS}$ obtained from the BESIII searches for
$\Sigma^+\to p+\mathrm{inv}$~\cite{BESIII:2023utd} and
$\Xi^0\to\Lambda+\mathrm{inv}$~\cite{BESIII:2026ztz}, compared with the
constraints from $K^+\to\pi^++\mathrm{inv}$~\cite{BNL-E949:2009dza,
NA62:2020pwi,NA62:2025upx}. The limits are evaluated for $m_a\simeq0$, with 
all other ALP coefficients set to zero, using \texttt{ALP-aca}~\cite{Alda:2025nsz}.}
\label{fig:baryons_eff}
\end{figure}

To determine the direct sensitivity to the flavour-changing
interactions, we vary $c_{ds}^{AP}$ and $c_{ds}^{VS}$ simultaneously,
while setting all other ALP coefficients to zero. The corresponding limits
are shown in Fig.~\ref{fig:baryons_eff}, together with those obtained
from the meson decay $K^+\to\pi^++\mathrm{inv}$. The baryon decays
provide direct sensitivity to the axial coefficient $c_{ds}^{AP}$,
which is not probed by pseudoscalar meson transitions. Their sensitivity
to the vector coefficient $c_{ds}^{VS}$ is currently weaker than that
of the kaon searches.

\subsection{Prompt ALP decays to muons}

The decay $\Sigma^+\to pa$ can also be probed when the ALP decays
promptly into a muon pair. This channel is restricted to
$2m_\mu\leq m_a\leq m_{\Sigma^+}-m_p$,
corresponding approximately to
$211\,\mathrm{MeV}\leq m_a\leq251\,\mathrm{MeV}$.
HyperCP reported an excess in $\Sigma^+\to p\mu^+\mu^-$ compatible
with a narrow state of mass
$m_{X^0}=214.3\pm0.5\,\mathrm{MeV}$~\cite{HyperCP:2005mvo}. The later
LHCb measurement did not confirm this excess~\cite{LHCb:2024fhb,
LHCb:2025evf}.

\begin{figure}
\centering
\includegraphics[width=0.48\textwidth]
{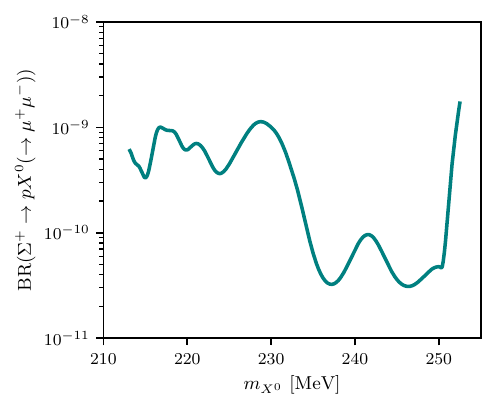}
\includegraphics[width=0.48\textwidth]
{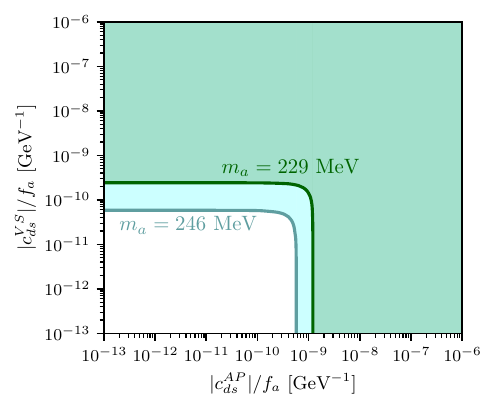}
\caption{Left: upper limit at the $90\%$ confidence level on
$\mathrm{BR}(\Sigma^+\to pX^0)\times
\mathrm{BR}(X^0\to\mu^+\mu^-)$ obtained from the reinterpretation of
the LHCb dimuon spectrum~\cite{LHCb:2024fhb,LHCb:2025evf}.
Right: $2\,\sigma$ exclusion limit on $c_{ds}^{AP}$ and $c_{ds}^{VS}$ assuming
$\mathrm{BR}(a\to\mu^+\mu^-)=1$. The results are shown for
$m_a=229\,\mathrm{MeV}$ and $m_a=246\,\mathrm{MeV}$, corresponding
respectively to the weakest and strongest constraints within the
accessible mass interval.}
\label{fig:BR_Sigmapmumu}
\end{figure}

We reinterpret the LHCb dimuon spectrum following the procedure of
Ref.~\cite{Abumusabh:2025zsr}. For each value of $m_{X^0}$, the fit
includes the SM contribution and a narrow signal smeared by the
reported dimuon mass resolution. The two normalisations are fitted
simultaneously, neglecting interference between the signal and the SM
amplitude. The signal yield is then increased until the fit reaches the
threshold corresponding to the $90\%$ confidence level. We convert the
resulting bound into a limit on
$\mathrm{BR}(\Sigma^+\to pX^0) \times
\mathrm{BR}(X^0\to\mu^+\mu^-)$
using a mass-independent single-event sensitivity $\alpha_{\rm eff}=4.56\times10^{-11}$, inferred from the combined
Run~2 signal yield and branching fraction reported in~\cite{LHCb:2025evf}.
The resulting limit is shown in the left panel of Fig.~\ref{fig:BR_Sigmapmumu}.

Assuming $\mathrm{BR}(a\to\mu^+\mu^-)=1$ allows this result to be expressed as 
a constraint on the invariant coefficients $c_{ds}^{AP}$ and $c_{ds}^{VS}$. 
The right panel of Fig.~\ref{fig:BR_Sigmapmumu} shows the $2\,\sigma$ exclusion 
limits for $m_a=229\,\mathrm{MeV}$ and $m_a=246\,\mathrm{MeV}$, which correspond
respectively to the weakest and strongest constraints within the accessible mass 
interval. Under the assumption of a unit muonic branching fraction, these limits 
are stronger than the BESIII constraints from invisible decays. 

\subsection{QED--DFSZ benchmark}

The interpretation above treats ALP production and decay as independent
and therefore requires an assumption about
$\mathrm{BR}(a\to\mu^+\mu^-)$. In a UV model, the coefficients
controlling these processes can instead be correlated. As an example,
we consider the QED--DFSZ model introduced in
Refs.~\cite{DiLuzio:2024jip,Alda:2025uwo}. At the matching scale
$\Lambda\simeq4\pi f_a$, the model predicts the couplings
\begin{eqnarray}
c_{u_R}^{ij} &=& -c_{d_R}^{ij} =
2\sin^2\beta\,\delta^{ij}, \qquad
c_{e_R}^{ij} = 2\cos^2\beta\,\delta^{ij}, \qquad c_B=-6, \label{eq:QED_DFSZ_matching} 
\\ 
c_{q_L}^{ij} &=& c_{\ell_L}^{ij}=0, \quad \qquad \qquad \qquad c_G = c_W=0. \nonumber
\end{eqnarray}

Renormalisation group evolution and matching across the heavy SM
thresholds generate effective flavour-changing ALP interactions through
the top quark coupling. For $\tan\beta=1$, the fermion coefficients in
Eq.~\eqref{eq:QED_DFSZ_matching} are universal up to a sign and generate
both ALP production in $\Sigma^+\to pa$ and $K^+\to\pi^+a$, as well as
the decay $a\to\mu^+\mu^-$. The production rate and the muonic
branching fraction can therefore be computed consistently within the
same model.

\begin{figure}
\centering
\includegraphics[width=0.7\textwidth]{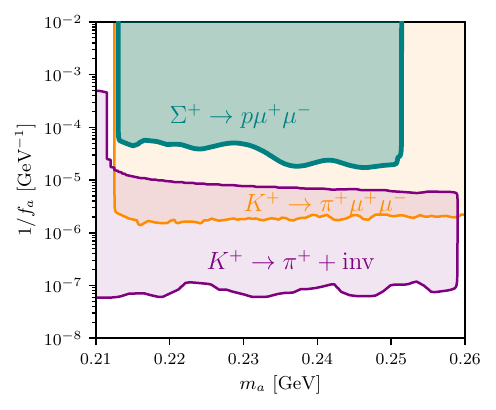}
\caption{$2\,\sigma$ exclusion limits on the QED--DFSZ model for $\tan\beta=1$. The
result from $\Sigma^+\to pa(\to\mu^+\mu^-)$ is compared with those from
$K^+\to\pi^+a(\to\mu^+\mu^-)$~\cite{NA482:2016sfh,
NA62:2022qes,NA62:2025upx} and
$K^+\to\pi^+a(\to\mathrm{inv})$~\cite{BNL-E949:2009dza,
NA62:2020pwi,NA62:2025upx}. The constraints are evaluated using
\texttt{ALP-aca}~\cite{Alda:2025nsz}.}
\label{fig:sda_decays}
\end{figure}

Figure~\ref{fig:sda_decays} compares the constraint derived from
$\Sigma^+\to pa(\to\mu^+\mu^-)$ with those obtained from visible and
invisible kaon decays. In the QED--DFSZ benchmark, the flavour-changing
vector and axial coefficients are correlated, $c_{ds}^{VS}=-c_{ds}^{AP}$, so all
three searches constrain the same underlying flavour-changing interaction. The present
$\Sigma^+\to p\mu^+\mu^-$ constraint is weaker by one order of magnitude than the $K^+\to\pi^+\mu^+\mu^-$ limits,
reflecting its current experimental reach. Nevertheless, it tests ALP
production in a distinct baryonic transition and, through the dimuon
signature, complements the kaon searches in visible and invisible final
states.

\section{Baryon pair channels above threshold}
\label{sec:baryonpairs}

For $m_a>2m_p$, the ALP can decay into baryon--antibaryon pairs. These
processes probe the flavour-conserving axial couplings discussed in
Sec.~\ref{sec:baryons} at timelike momentum transfer of order a few
GeV. We evaluate the exclusive decay widths using the continuation
defined in Eq.~\eqref{eq:baryon_source_dressing}.

To assess the relative importance of the exclusive baryon channels, we
compare their sum with the inclusive perturbative QCD estimate
\begin{equation}
\Gamma(a\to\mathrm{hadrons})
\simeq
\Gamma(a\to gg)
+
\sum_{q=u,d,s}\Gamma(a\to q\bar q).
\label{eq:pQCD_total}
\end{equation}
For $N_f=3$, the decay width into gluons is
\begin{equation}
\Gamma(a\to gg) = (N_c^2-1)\frac{\alpha_s^2(m_a)m_a^3}{(4\pi)^3f_a^2}
\left[ 1+83\frac{\alpha_s(m_a)}{4\pi} \right] |c_{gg}|^2 .
\end{equation}
which includes the one-loop QCD correction~\cite{Spira:1995rr}. The contribution of quark loops is accounted for through the effective coupling~\cite{Bauer:2017ris},
\begin{equation}
c_{gg} = c_{GP} +\sum_{q=u,d,s} c_{qq}^{AP} B_1(4m_q^2/m_a^2) \xrightarrow{m_a\gg m_s} c_{GP}+\langle \bm{c_q^{AP}}\rangle=c_{GA}.
\end{equation}
The high-mass limit follows from the singlet axial Ward identity.
After integrating the derivative ALP--quark interaction by parts, the
divergence of the axial current contains both a quark-mass term and the
gluonic anomaly. For $m_a^2\gg m_q^2$, the mass-dependent contribution
is suppressed, whereas the anomalous contribution survives, corresponding
to $B_1(4m_q^2/m_a^2)\to1$. The limiting coefficient is therefore the
field-redefinition-invariant anomalous combination
$c_{GP}+\langle\bm{c_q^{AP}}\rangle=c_{GA}$.
In particular, for a purely pseudoscalar ALP--quark interaction with
$c_G=0$ and $\bm{c_q^A}=0$, this relation implies $c_{GA}=0$ and hence the
decoupling of the gluonic amplitude in the high-mass limit~\cite{Alda:2024cxn}.
The decay width into a quark-antiquark pair is
\begin{equation}
\Gamma(a\to q\bar q) = N_c\frac{m_a\overline m_q^{\,2}(m_a)}{2\pi f_a^2}
\left[1+\frac{17}{3}\frac{\alpha_s(m_a)}{\pi}\right] |c_{qq}^{AP}|^2
\sqrt{1-\frac{4m_q^2}{m_a^2}},
\end{equation}
where $\overline{m}_q(m_a)$ is the running quark mass in the $\overline{\mathrm{MS}}$ 
scheme. The term proportional to $\alpha_s(m_a)$ incorporates the one-loop QCD
correction~\cite{Chetyrkin:1994kg}.
We neglect the gluon-induced contribution to the effective ALP--quark
coupling, which starts at $\mathcal O(\alpha_s^2)$ in the amplitude and
is numerically irrelevant for the comparison considered here.

Equation~\eqref{eq:pQCD_total} relies on quark--hadron duality: at
sufficiently large $m_a$, the inclusive hadronic width, summed over all
kinematically accessible final states, is expected to approach its
perturbative quark--gluon estimate. In the few-GeV region, threshold and
resonance effects can produce sizeable deviations, so we use
Eq.~\eqref{eq:pQCD_total} only as a reference estimate for the inclusive
rate~\cite{Poggio:1975af,Shifman:2000jv,Chibisov:1996wf}.
The inclusive rate contains all kinematically accessible hadronic final
states, including states with higher multiplicity and hadrons not
included in the baryon octet.

For a diagonal axial coupling, the exclusive decay width is
\begin{equation}
\Gamma(a\to B\overline B)
=
\frac{m_am_B^2}{2\pi f_a^2}
|c_{BB}^A(m_a)|^2
\sqrt{1-\frac{4m_B^2}{m_a^2}}.
\label{eq:ALP_baryon_pair_width}
\end{equation}
This exclusive prediction inherits the form factor and matching scale
dependence of the continuation introduced in
Sec.~\ref{sec:baryon_GeV_continuation}.

\begin{figure}
\centering
\includegraphics[width=0.49\textwidth]
{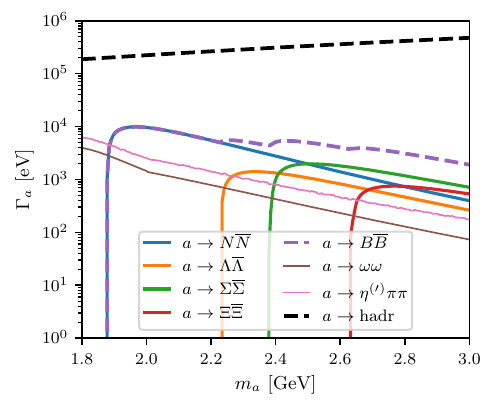}
\includegraphics[width=0.49\textwidth]
{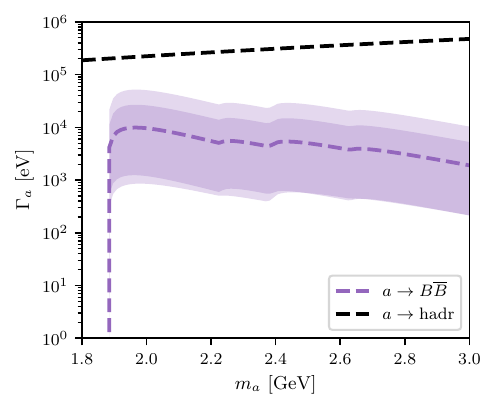}
\caption{ALP decay widths into baryon pairs in the gluophilic benchmark
of Eq.~\eqref{eq:gluophilic_benchmark}, with $f_a=1/(16\pi^2)\,\mathrm{TeV}$ and $\mu_\star = 1.9\,\mathrm{GeV}$. Left: individual widths into
nucleons, $\Lambda$ baryons, $\Sigma$ baryons, and $\Xi$ baryons for
$\widetilde S_B=0.025$. The sum of the exclusive baryon widths is compared
with the inclusive pQCD estimate and with the exclusive
meson channels $a\to\omega\omega$ and $a\to\eta^{(\prime)}\pi\pi$. Right: variation of the sum of
the exclusive baryon widths for different values of $\widetilde{S}_B$: in darker purple in the na\"ive prescription and lighter in the conservative prescription.}
\label{fig:dw_baryons}
\end{figure}

The left panel of Fig.~\ref{fig:dw_baryons} shows the individual
baryon pair widths for $\widetilde S_B=0.025$, their sum, and the inclusive
width in Eq.~\eqref{eq:pQCD_total}. The exclusive mesonic channels commonly considered in the literature are also shown for comparison.

The inclusive rate contains hadronic states with higher
multiplicity, as well as heavier hadrons that are absent from the
exclusive sum. Nevertheless, the left panel of
Fig.~\ref{fig:dw_baryons} shows that baryon pair channels provide a
non-negligible contribution in the mass region where they are
kinematically accessible.
Their summed width is comparable to the exclusive meson channels, such as $a\to \eta^{(\prime)}\pi\pi$ and $a\to\omega\omega$, commonly considered in the literature~\cite{Aloni:2018vki,Ovchynnikov:2025gpx,Balkin:2025enj,Bai:2025fvl}. Near $m_a\simeq2\,\mathrm{GeV}$, the
nucleon-pair width reaches approximately $5\%$ of the perturbative
inclusive estimate. At larger masses, the sum of the baryon-pair widths
remains at the order $0.1-5\%$ level relative to the same estimate, while
$\Sigma$ and $\Xi$ pairs become more important than nucleon pairs
shortly above their respective thresholds. 

The baryonic decay widths are affected by the uncertainty associated to $\widetilde{S}_B$. The total baryonic decay width for several ranges of this coefficient is presented in the right panel of Fig.~\ref{fig:dw_baryons}. In the na\"ive prescription $\widetilde{S}_B = 0.025\pm0.031$, baryon channels may represent between 0.6\% and 12\% of the inclusive pQCD estimate at $m_a=2\,\mathrm{GeV}$. 

\begin{figure}
    \centering
    \includegraphics[width=0.495\textwidth]{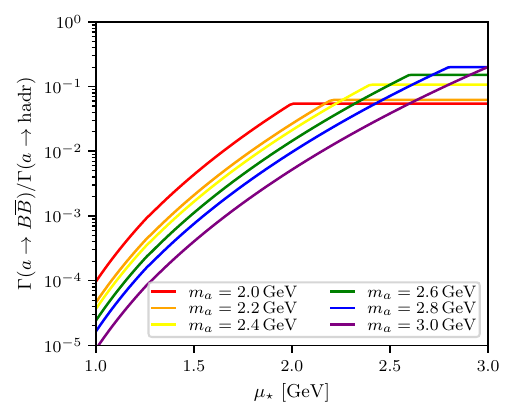} \includegraphics[width=0.495\textwidth]{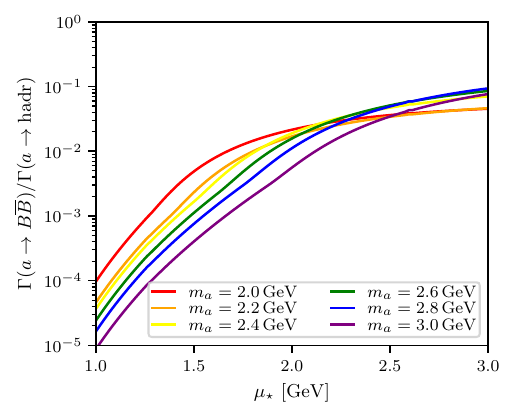}
    \caption{Dependence on $\mu_\star$ of the baryonic branching ratios for an ALP in the gluophilic scenario with $f_a=1/(16\pi^2)\,\mathrm{TeV}$ and $\widetilde{S}=0.025$. Left: using the piecewise form factors of Eq.~\eqref{eq:baryon_continuation_profile}. Right: interpolating the two regimes with a cubic spline.}
    \label{fig:mus_dependence}
\end{figure}

The crossover scale $\mu_\star$ is another major source of uncertainty of this
continuation. It is not fixed by the asymptotic scaling laws, but
parametrises where the low-energy description is replaced by the
hard-scattering envelope. As phenomenological guidance, the onset of
approximately perturbative scaling in spacelike nucleon form factors has 
been estimated at $-q^2\simeq 5-6\,\mathrm{GeV}^2$ \cite{Stoler:1993yk}.
This motivates exploring $\mu_\star\gtrsim2\,\mathrm{GeV}$, but it does not 
determine the matching point for the timelike axial form factors considered 
here. For the sharp profile in Eq.~\eqref{eq:baryon_continuation_profile}
(left panel of Fig.~\ref{fig:mus_dependence}), varying $\mu_\star$
over $1-3\,\mathrm{GeV}$ changes the baryonic width by
up to two orders of magnitude for $m_a=2\,\mathrm{GeV}$ and three
orders of magnitude for $m_a=3\,\mathrm{GeV}$. This large variation
also reflects the abrupt matching ansatz. Replacing it with a cubic
spline in the intermediate region, as shown in the right panel of Fig.~\ref{fig:mus_dependence}, reduces the variation
to below a factor of ten for $2\,\mathrm{GeV}\leq\mu_\star\leq3\,\mathrm{GeV}$, 
for all ALP masses displayed.
With $\mu_\star \gtrsim 2\,\mathrm{GeV}$ and a soft interpolation to the pQCD-like regime,
baryonic channels constitute $\sim 1\%-10\%$ of the hadronic decay width for a large range of ALP masses.
Comparable form-factor and 
matching ambiguities also affect the mesonic estimates in
Refs.~\cite{Aloni:2018vki,Ovchynnikov:2025gpx,Balkin:2025enj}.

A recent symmetry-constrained study of few-GeV hadronic 
decays~\cite{Kryshtal:2026had} based on \textsc{Pythia} provides a complementary
event-level perspective. For its fermion universal ALP benchmark, only
the $p\bar p$ and $n\bar n$ widths are retained as separately
calculated baryonic channels, while the partonic $gg$ and $s\bar s$
rates determine the mixture of fragmentation sources rather than
additional contributions to the hadronic width. Hyperon pair channels
are not resolved through separate exclusive amplitudes.
The analysis therefore reinforces the conclusion that the available
exclusive calculations do not saturate the inclusive hadronic width
and that the residual contribution must be assigned to additional
multihadron final states. It also demonstrates that unmodified
\textsc{Pythia} should be used with caution in this mass range.

\section{Conclusions}\label{sec:conclusions}

We have developed a field redefinition invariant description of hadronic amplitudes 
outside the odd-intrinsic-parity WZW sector at $\mathcal O(1/f_a)$.  In contrast to 
existing field redefinition invariant formulations of strong interaction hadronic 
amplitudes, which have so far been restricted to flavour conserving ALP--quark 
interactions, our construction retains general non diagonal flavour structures.
Starting from 
the general CP-conserving ALP--QCD Lagrangian, with both flavour conserving and 
flavour changing interactions, we identified the invariant combinations of partonic
couplings and matched the $aG\widetilde G$ interaction through the anomalous $U(1)_A$
sector of the $U(3)$ chiral theory. After diagonalising the SM pseudoscalar sector,
the LSZ construction determines the axial field redefinition that removes the
ALP--meson mixing amplitudes. In the resulting no-mixing parametrisation, the 
direct ALP couplings are expressed exclusively in terms of invariant quantities. 
Field redefinition invariance is therefore manifest at the level of individual 
vertices, without requiring cancellations between basis dependent direct and mixing
contributions.

The baryon octet provides a substantive application of this framework. We derived 
the general vector and axial ALP--baryon couplings, including flavour changing 
transitions. A central new result is the identification of the second topological 
singlet structure of the general $U(3)$ baryon Lagrangian as an independent 
interaction of a dynamical ALP with the baryon octet. This structure was 
originally introduced for the derivative of the external QCD vacuum angle in 
Ref.~\cite{Bass:1999is,Borasoy:1999nd,Borasoy:2000xj}, but was not retained as a dynamical ALP--baryon 
interaction. To our knowledge, it is included in this context for the first time 
here. 
We have also performed an extraction of the corresponding coefficient based on lattice QCD inputs.

The resulting flavour changing hyperon amplitudes probe axial $s\to da$ interactions 
that are inaccessible in pseudoscalar-to-pseudoscalar meson decays. Present BESIII 
searches therefore provide information complementary to that obtained from kaon
decays. Within the restricted dimuon mass window and under the assumption $\mathrm{BR} 
(a\to\mu^+\mu^-)=1$, our reinterpretation of the LHCb spectrum gives stronger 
constraints than the current invisible hyperon searches. The QED--DFSZ benchmark 
demonstrates that running and matching can generate sizeable flavour changing 
interactions at low energies from flavour-diagonal ultraviolet boundary conditions, 
and that the resulting flavour changing observables can provide some of the
strongest constraints on the ultraviolet parameter space.

Previous extensions of GeV scale ALP phenomenology have primarily focused on enlarging 
the mesonic spectrum. Above the baryon-pair threshold, our form factor continuation 
shows that baryon--antibaryon final states can constitute a relevant component of the 
exclusive hadronic ALP width in the gluophilic benchmark considered here. The
summed baryonic width is comparable to, and over part of the displayed mass range 
larger than, the individual $a\to\omega\omega$ and $a\to\eta^{(\prime)}\pi\pi$ widths commonly 
considered in the literature. Nucleon pairs can reach approximately $5\%$ of the 
inclusive perturbative estimate near $2\,\mathrm{GeV}$, while the total exclusive
baryonic contribution remains at the $0.1$--$5\%$ level at larger masses. Most of the 
inclusive width must therefore reside in additional hadronic final states. Both the 
baryonic prediction and its comparison with exclusive mesonic channels carry model 
dependence associated with the form factor continuation and matching scale. The baryonic 
width also depends strongly on the independent coefficient $\widetilde S_B$ and can 
be substantially suppressed or enhanced within its prescribed range.

The present construction applies at $\mathcal O(1/f_a)$, outside the odd intrinsic 
parity WZW sector, at leading chiral order, for perturbative ALP--meson mixing with 
$f_\pi/f_a\ll1$, and away from pseudoscalar degeneracies. A broader description will 
require combining this framework with the invariant treatments of WZW-mediated
amplitudes, incorporating higher order chiral corrections and heavier resonances, 
and treating the near-degenerate ALP--meson system beyond perturbation theory. Within 
its stated domain, the framework provides a general interface between ultraviolet ALP 
couplings and hadronic amplitudes without requiring the gluonic operator to be removed 
by a preliminary chiral rotation.

\section*{Acknowledgements}
We acknowledge partial financial support by the European Union’s Horizon 2020 
research and innovation programme under the Marie Sk{\l}odowska-Curie grant agreement 
N$^o$. 101086085-ASYMMETRY. We thank Luca Di Luzio and Luca Merlo for helpful 
discussions and Gabriele Levati for clarifications about Ref.~\cite{DiLuzio:2023cuk}.
The authors utilized GPT 5.6 to refine the English phrasing and language flow of 
the manuscript, and for analytical and numerical checks of the asymptotic behaviour of the form factor  $G_{\widetilde{G}}(q^2)$ and the extraction of $\widetilde{S}_B$. The authors reviewed and edited all outputs and take full 
responsibility for the contents of the manuscript.

\clearpage
\onecolumngrid
\appendix

\section{Expansion in the number of pseudoscalar fields}\label{sec:Phi_expansion}

In this appendix, we provide a method for expanding the CCWZ building
blocks to arbitrary order in the pseudoscalar fields. For compactness,
we introduce the dimensionless matrix
\begin{equation}
\bm{\varphi} = \frac{\bm{\Phi}}{2f_\pi}.
\end{equation}
Let us define the linear operators for left- and right-multiplication,
\begin{equation}
    \mathbb{L}_{\bm{\varphi}}[\bm{A}] = \bm{\varphi}\bm{A},\qquad\mathbb{R}_{\bm{\varphi}}[\bm{A}]= \bm{A}\bm{\varphi},
\end{equation}
which commute, $[\mathbb{L}_{\bm{\varphi}},\mathbb{R}_{\bm{\varphi}}] = 0$, since $\mathbb{L}_{\bm{\varphi}} \mathbb{R}_{\bm{\varphi}}[\bm{A}] = \mathbb{R}_{\bm{\varphi}} \mathbb{L}_{\bm{\varphi}}[\bm{A}] = \bm{\varphi}\bm{A}\bm{\varphi}$. It follows that
\begin{equation}\begin{split}
    \bm{u}\bm{A} = \exp(i\bm{\varphi})\bm{A} = \sum_k \frac{i^k}{k! }\bm{\varphi}^k\bm{A} =  \sum_k \frac{i^k}{k! }\mathbb{L}_{\bm{\varphi}}^k[\bm{A}] \equiv \exp(i\,\mathbb{L}_{\bm{\varphi}})[\bm{A}],\\
    \bm{u}^\dagger\bm{A} = \exp(-i\,\mathbb{L}_{\bm{\varphi}})[\bm{A}],\qquad \bm{A}\bm{u} = \exp(i\,\mathbb{R}_{\bm{\varphi}})[\bm{A}],\qquad  \bm{A}\bm{u}^\dagger = \exp(-i\,\mathbb{R}_{\bm{\varphi}})[\bm{A}].
\end{split}
\end{equation}
Using the commutativity of the operators,
\begin{equation}\begin{split}
    \bm{u}\bm{A}\bm{u} = \exp\bigg(i(\mathbb{L}_{\bm{\varphi}}+\mathbb{R}_{\bm{\varphi}})\bigg)[\bm{A}],\qquad\bm{u}^\dagger\bm{A}\bm{u}^\dagger = \exp\bigg(-i(\mathbb{L}_{\bm{\varphi}}+\mathbb{R}_{\bm{\varphi}})\bigg)[\bm{A}],\\
    \bm{u}\bm{A}\bm{u}^\dagger = \exp\bigg(i(\mathbb{L}_{\bm{\varphi}}-\mathbb{R}_{\bm{\varphi}})\bigg)[\bm{A}],\qquad\bm{u}^\dagger\bm{A}\bm{u} = \exp\bigg(-i(\mathbb{L}_{\bm{\varphi}}-\mathbb{R}_{\bm{\varphi}})\bigg)[\bm{A}].
\end{split}\end{equation}
It is useful to introduce the operators
\begin{equation}
\begin{aligned}
\mathbb A_{\bm{\varphi}}[\bm A]
&\equiv
(\mathbb L_{\bm{\varphi}}+\mathbb R_{\bm{\varphi}})[\bm A]
=
\{\bm{\varphi},\bm A\},
\\
\mathbb C_{\bm{\varphi}}[\bm A]
&\equiv
(\mathbb L_{\bm{\varphi}}-\mathbb R_{\bm{\varphi}})[\bm A]
=
[\bm{\varphi},\bm A].
\end{aligned}
\end{equation}
Consequently, $\mathbb A_{\bm{\varphi}}^k[\bm A]$ and
$\mathbb C_{\bm{\varphi}}^k[\bm A]$ generate $k$ nested
anticommutators and commutators, respectively.

With these properties, we can obtain the expansion of $\bm{\chi_\pm}$ in the number of pseudoscalar mesons. If we write $\bm{\chi} = \bm{\chi_S} + i \bm{\chi_P}$ with $\bm{\chi_{S,P}}$ Hermitian, then, recalling the definition,
\begin{equation}
\bm{\chi_\pm} = \bm{u}^\dagger \bm{\chi}\bm{u}^\dagger \pm \bm{u} \bm{\chi}^\dagger\bm{u} \nonumber
\end{equation}
one obtains
\begin{equation}\begin{split}
    \bm{\chi_+} &=2 \cos(\mathbb{A}_{\bm{\varphi}})[\bm{\chi_S}] + 2 \sin(\mathbb{A}_{\bm{\varphi}})[\bm{\chi_P}] = 2\left(\bm{\chi_S} + \left\{\bm{\varphi},\bm{\chi_P}\right\} - \frac{1}{2} 
    \left\{\bm{\varphi},\left\{\bm{\varphi},\bm{\chi_S}\right\}\right\} + \cdots \right) \\
    \bm{\chi_-} &= 2i \cos(\mathbb{A}_{\bm{\varphi}})[\bm{\chi_P}]-2i \sin(\mathbb{A}_{\bm{\varphi}})[\bm{\chi_S}] =
    2i\left(\bm{\chi_P} - \left\{\bm{\varphi},\bm{\chi_S}\right\} - \frac{1}{2} 
    \left\{\bm{\varphi},\left\{\bm{\varphi},\bm{\chi_P}\right\}\right\} + \cdots \right).
    \nonumber
\end{split}\end{equation}
In particular, the expansion of $\bm{\chi_+}$ retaining terms up to two fields (pseudoscalar mesons and/or ALP) is
\begin{equation}\begin{split}
    \bm{\chi_+} &\supset 4B_0 \bm{m_q} + 4i B_0[\bm{c_q^S},\bm{m_q}]\frac{a}{f_a} + \frac{2B_0}{f_\pi}\{\bm{\Phi},\{\bm{c_q^P},\bm{m_q}\}\}\frac{a}{f_a}-\frac{B_0}{2f_\pi^2} \{\bm{\Phi},\{\bm{\Phi},\bm{m_q}\}\}.
\end{split}
\end{equation}
We can also compute the general expansion of the compensator $\bm{h}$. If we define $\bm{h} = \exp[\frac{ia}{f_a}(\bm{\delta} + \bm{\zeta})]$ and perform an infinitesimal chiral rotation of $\bm{u}$:
\begin{equation}\begin{split}
    \bm{u} &\to \bm{g_R}\bm{u} \bm{h}^\dagger \approx\bm{u} + \frac{ia}{f_a}\bigg([\bm{\delta},\bm{u}] + \bm{\kappa}\bm{u} - \bm{u}\bm{\zeta}\bigg),\\
    \bm{u} &\to \bm{h}\bm{u} \bm{g_L}^\dagger \approx\bm{u} + \frac{ia}{f_a}\bigg([\bm{\delta},\bm{u}] + \bm{\zeta}\bm{u} + \bm{u}\bm{\kappa}\bigg).
\end{split}\end{equation}
The condition for equality of the two expressions is $\bm{\zeta}\bm{u} + \bm{u}\bm{\zeta} = \bm{\kappa}\bm{u} - \bm{u}\bm{\kappa}$. Multiplying both sides on the right by $\bm{u}^\dagger$,
\begin{equation}
    \left[1+\exp(-i\,\mathbb{C}_{\bm{\varphi}})\right][\bm{\zeta}] = \left[\exp(-i\,\mathbb{C}_{\bm{\varphi}})-1\right][\bm{\kappa}],
\end{equation}
from which we can solve for $\bm{\zeta}$,
\begin{equation}
    \bm{\zeta} = \frac{\exp(-i\,\mathbb{C}_{\bm{\varphi}})-1}{\exp(-i\,\mathbb{C}_{\bm{\varphi}})+1}[\bm{\kappa}] = -i \tan\left(\frac{\mathbb{C}_{\bm{\varphi}}}{2}\right)[\bm{\kappa}],
\end{equation}
and finally inserting in the compensator,
\begin{equation}
    \bm{h} = \bm{1} + \frac{a}{f_a}\left[i\bm{\delta} + \tan\left(\frac{\mathbb{C}_{\bm{\varphi}}}{2}\right)[\bm{\kappa}] \right] =  \bm{1}+ \frac{a}{f_a}\left(i\bm{\delta} +\frac{1}{4f_\pi}[\bm{\Phi},\bm{\kappa}]+\frac{1}{192 f_\pi^3}[\bm{\Phi},[\bm{\Phi},[\bm{\Phi}, \bm{\kappa}]]] + \ldots\right).
\end{equation}
For the vielbein and the connection, we also require the expansion of 
$\partial_\mu \bm{u}$. Since $[\bm{\Phi},\partial_\mu \bm{\Phi}] \neq 0$, 
the derivative of a matrix exponential is
\begin{equation}
    \partial_\mu \bm{u} = \partial_\mu \exp(i\,\bm{\varphi}) = \int_0^1ds\ \exp[i(1-s)\bm{\varphi}](i\,\partial_\mu \bm{\varphi}) \exp(is\bm{\varphi}).
\end{equation}
By using the multiplication operators,
\begin{equation}
    \partial_\mu \bm{u} =i\, \exp(i\,\mathbb{L}_{\bm{\varphi}}) \int_0^1 ds\ \exp(-is\,\mathbb{C}_{\bm{\varphi}})[\partial_\mu\bm{\varphi}] = - \exp(i\mathbb{L}_{\bm{\varphi}}) \frac{\exp(-i\,\mathbb{C}_{\bm{\varphi}})-1}{\mathbb{C}_{\bm{\varphi}}}[\partial_\mu\bm{\varphi}],
\end{equation}
and consequently
\begin{equation}\begin{split}
    \bm{u}^\dagger \partial_\mu \bm{u} = -\frac{\exp(-i\,\mathbb{C}_{\bm{\varphi}})-1}{\mathbb{C}_{\bm{\varphi}}}[\partial_\mu\bm{\varphi}],\qquad
    \bm{u} \partial_\mu \bm{u}^\dagger = -\frac{\exp(i\,\mathbb{C}_{\bm{\varphi}})-1}{\mathbb{C}_{\bm{\varphi}}}[\partial_\mu\bm{\varphi}].
\end{split}\end{equation}

We obtain the following expressions for the vielbein and the connection
\begin{equation}\begin{split}
    \bm{u_\mu} &= -2\frac{\sin(\mathbb{C}_{\bm{\varphi}})}{\mathbb{C}_{\bm{\varphi}}}[\partial_\mu \bm{\varphi}] + 2\frac{\partial_\mu a}{f_a}\cos(\mathbb{C}_{\bm{\varphi}})[\bm{c_q^A}]- 2i\frac{\partial_\mu a}{f_a}\sin(\mathbb{C}_{\bm{\varphi}})[\bm{c_q^V}],\\
    \bm{\Gamma_\mu} &= \frac{1-\cos(\mathbb{C}_{\bm{\varphi}})}{\mathbb{C}_{\bm{\varphi}}}[\partial_\mu \bm{\varphi}] - \frac{\partial_\mu a}{f_a}\sin(\mathbb{C}_{\bm{\varphi}})[\bm{c_q^A}]-i\frac{\partial_\mu a}{f_a}\cos(\mathbb{C}_{\bm{\varphi}})[\bm{c_q^V}],
\end{split}
\end{equation}
and the terms of the expansion with the lowest number of pNGB fields are:
\begin{equation}\begin{split}
    \bm{u_\mu} &\supset -\frac{1}{f_\pi}\partial_\mu \bm{\Phi}+ 2\frac{\partial_\mu a}{f_a}\bm{c_q^A}-i\frac{\partial_\mu a}{f_\pi f_a}[\bm{\Phi},\bm{c_q^V}],\\
    \bm{\Gamma_\mu} &\supset-i \frac{\partial_\mu a}{f_a}\bm{c_q^V}+\frac{1}{8f_\pi^2}[\bm{\Phi},\partial_\mu\bm{\Phi}]-\frac{\partial_\mu a}{2 f_\pi f_a}[\bm{\Phi},\bm{c_q^A}].
\end{split}\end{equation}

Finally, we derive the transformation law for the nonet of pseudoscalar mesons $\bm\varphi$. For this purpose, it is useful to work with $\bm U = \bm u^2 = \exp(2i\bm\varphi)$, which transforms as $\bm U \to  \bm{g_R} \bm U \bm{g_L}^\dagger = \bm U + \delta \bm U$. Writing this explicitly,
\begin{equation}
    \delta\bm{U} = i\frac{a}{f_a}(\bm \delta + \bm \kappa)\bm U + i\frac{a}{f_a} \bm{U}(\bm\kappa - \bm \delta) = i\frac{a}{f_a} \left[(1-e^{2i\mathbb C_{\bm\varphi}})[\bm\delta]
 +(1+e^{2i\mathbb C_{\bm\varphi}})[\bm\kappa]\right] \bm U.
\end{equation}
On the other hand, under an infinitesimal variation $\bm \varphi \to \bm{\varphi'} = \bm\varphi + \delta \bm\varphi$, the variation of $\bm U$ is
\begin{equation}
    \delta  \bm U = 2i \int_0^1 ds\ e^{2is\bm\varphi}(\delta\bm\varphi)e^{2i(1-s)\bm\varphi} = \frac{e^{2i\mathbb{C}_{\bm\varphi}} -1}{\mathbb{C}_{\bm\varphi}}[\delta\bm\varphi]\bm U.
\end{equation}
Combining these two expressions gives
\begin{equation}\begin{split}
\delta\bm\varphi =& i\frac{a}{f_a} \frac{\mathbb{C}_{\bm\varphi}}{e^{2i\mathbb{C}_{\bm\varphi}} -1}\left[(1-e^{2i\mathbb C_{\bm\varphi}})[\bm\delta]
 +(1+e^{2i\mathbb C_{\bm\varphi}})[\bm\kappa]\right]\\
 =& i\frac{a}{f_a}[\bm\delta,\bm\varphi] + \frac{a}{f_a} \mathbb{C}_{\bm\varphi} \cot \mathbb{C}_{\bm\varphi}[\bm{\kappa}].
\end{split}\end{equation}

\section{Completeness relations for the pseudoscalar meson basis}\label{sec:app_completeness}
We begin by noting that, owing to the orthogonality relations $\langle \bm{\lambda_i} \bm{\lambda_j^\dagger}\rangle
= 2\delta_{ij}$, any $N_f\times N_f$ matrix can be projected as
\begin{equation}
    \bm{A} = \frac{1}{2} \sum_i \langle\bm{\lambda_i}^\dagger\bm{A}\rangle \bm{\lambda_i},
\end{equation}
and inserting another matrix $\bm{B}$ and taking the trace,
\begin{equation}
    \langle\bm{A} \bm{B}\rangle = \frac{1}{2} \sum_i \langle\bm{\lambda_i}^\dagger\bm{A}\rangle \langle\bm{\lambda_i} \bm{B}\rangle.
\end{equation}

To resum expressions involving the meson masses $m_i^2$, we define the linear operator
\begin{equation}
    \mathbb{M}(\bm{A}) = B_0\{\bm{m_q},\bm{A}\} + \frac{m_0^2}{N_f}\langle \bm{A}\rangle \bm{1},
\end{equation}
which satisfies the eigenvalue equation
\begin{equation}
    \mathbb{M}(\bm{\lambda_i}) = m_i^2 \bm{\lambda_i}.
\end{equation}
Then, for any matrix $\bm{A}$ we have
\begin{equation}
    \langle\bm{A}\bm{\lambda_i}^\dagger\rangle m_i^2 = \langle\bm{A}\, \mathbb{M}(\bm{\lambda_i})^\dagger\rangle = \langle\mathbb{M}(\bm{A})\bm{\lambda_i}^\dagger\rangle,
\end{equation}
where the last equality follows because $\mathbb{M}$ is a self-adjoint operator.

In particular, in the computation of the LSZ factors $w_i^{(a)}$,
\begin{equation}\begin{split}
    w_i^{(a)} &= \frac{K_{i\hat{a}}m_a^2 -M_{i\hat{a}}^2}{m_i^2-m_a^2} = -K_{i\hat{a}} + \frac{K_{i\hat{a}}m_i^2 - M_{i\hat{a}}^2}{m_i^2-m_a^2} = \\&=\frac{f_\pi}{f_a}\bigg(\langle\bm{c_q^A}\bm{\lambda_i}^\dagger\rangle + \frac{-\langle\bm{c_q^A}\bm{\lambda_i}^\dagger\rangle m_i^2 +B_0\langle\{\bm{c_q^P},\bm{m_q}\}\bm{\lambda_i}^\dagger\rangle - \frac{m_0^2}{N_f}c_G\langle\bm{\lambda_i}^\dagger\rangle }{m_i^2-m_a^2} \bigg)\\
    &= \frac{f_\pi}{f_a}\bigg(\langle\bm{c_q^A}\bm{\lambda_i}^\dagger\rangle + \frac{-\langle\mathbb{M}(\bm{c_q^A})\bm{\lambda_i}^\dagger\rangle +B_0\langle\{\bm{c_q^P},\bm{m_q}\}\bm{\lambda_i}^\dagger\rangle - \frac{m_0^2}{N_f}c_G\langle\bm{\lambda_i}^\dagger\rangle }{m_i^2-m_a^2} \bigg)\\
    &= \frac{f_\pi}{f_a}\bigg(\langle\bm{c_q^A}\bm{\lambda_i}^\dagger\rangle + \frac{-B_0\langle\{\bm{m_q},\bm{c_q^A}\}\bm{\lambda_i}^\dagger\rangle -\frac{m_0^2}{N_f}\langle\bm{c_q^A}\rangle\langle\bm{\lambda_i}^\dagger\rangle+B_0\langle\{\bm{c_q^P},\bm{m_q}\}\bm{\lambda_i}^\dagger\rangle - \frac{m_0^2}{N_f}c_G\langle\bm{\lambda_i}^\dagger\rangle }{m_i^2-m_a^2} \bigg)\\
    &= \frac{f_\pi}{f_a}\bigg(\langle\bm{c_q^A}\bm{\lambda_i}^\dagger\rangle + \frac{\langle\bm{\mathcal{C}}\bm{\lambda_i}^\dagger\rangle  }{m_i^2-m_a^2} \bigg),
\end{split}
\end{equation}
where we have defined
\begin{equation}
    \bm{\mathcal{C}} = B_0 \{\bm{m_q},\bm{c_q^P}\} - \frac{m_0^2}{N_f}c_G\bm{1} - \mathbb{M}(\bm{c_q^A}) = -B_0 \{\bm{m_q},\bm{c_q^{AP}}\} - \frac{m_0^2}{N_f}c_{GA}\bm{1}.
\end{equation}

We also compute sums of the type
\begin{equation}
    \sum_i \frac{\langle\bm{\lambda_i}^\dagger \bm{A}\rangle \langle \bm{\lambda_i} \bm{B}\rangle }{m_i^2}.
\end{equation}

The inverse of $\mathbb{M}$, using the spectral decomposition, is
\begin{equation}
    \mathbb{M}^{-1}(\bm{A}) = \frac{1}{2}\sum_i \frac{\langle \bm{A}\bm{\lambda_i}^\dagger\rangle}{m_i^2}\bm{\lambda_i},
\end{equation}
so the sum is given by
\begin{equation}
    \sum_i \frac{\langle\bm{\lambda_i}^\dagger \bm{A}\rangle \langle \bm{\lambda_i} \bm{B}\rangle }{m_i^2} = 2 \langle\mathbb{M}^{-1}(\bm{A}) \bm{B}\rangle =  2 \langle\mathbb{M}^{-1}(\bm{B}) \bm{A}\rangle.
\end{equation}

To obtain $\mathbb M^{-1}(\bm A)$ without using its spectral decomposition, we exploit 
the fact that $\bm{m_q}$ is diagonal. For $\bm A=\mathbb M(\bm Y)$,
\begin{equation}
    (\bm{A})_{qq'} = \big(\mathbb{M}(\bm{Y})\big)_{qq'} = B_0 (m_q+m_{q'}) (\bm{Y})_{qq'} +\frac{m_0^2}{N_f}\langle\bm{Y}\rangle\delta_{qq'}.
\end{equation}
For the off-diagonal elements, the relation can be inverted directly,
\begin{equation}
    \big(\mathbb{M}^{-1}(\bm{A})\big)_{qq'} = (\bm{Y})_{qq'} = \frac{(\bm{A})_{qq'}}{B_0 (m_q+m_{q'})},\qquad q \neq q'.
\end{equation}
For the diagonal elements (with no sum over repeated indices),
\begin{equation}
    \big(\mathbb{M}^{-1}(\bm{A})\big)_{qq} = (\bm{Y})_{qq} = \frac{1}{2B_0 m_q} \left[(\bm{A})_{qq} -\frac{\langle \bm{A} \bm{m_q}^{-1}\rangle/B_0}{\frac{2N_f}{m_0^2} + \langle\bm{m_q}^{-1}\rangle/B_0} \right].
\end{equation}
ALP-meson mixing effects in the massless ALP limit are encoded in $\mathbb{M}^{-1}(\bm{\mathcal{C}})$, with $\bm{\mathcal{C}}$ defined in Eq.~\eqref{eq:mixing_coeff}, obtaining
\begin{equation}
    \mathbb{M}^{-1}(\bm{\mathcal{C}})= -\bm{c_q^{AP}}- c_{GP}\frac{\, \bm{m_q}^{-1}/B_0}{\frac{2N_f}{m_0^2} + \langle\bm{m_q}^{-1}\rangle/B_0}.\label{eq:minv_low_ma}
\end{equation}
For $m_0^2 \gg B_0 m_s \gg B_0 m_d$, 
\begin{equation}
\begin{split}
    \mathbb{M}^{-1}(\bm{\mathcal{C}}) &\approx  -\bm{c_q^{AP}}-  c_{GP}\frac{\, \bm{m_q}^{-1}}{ \langle\bm{m_q}^{-1}\rangle} \approx  -\bm{c_q^{AP}} -  c_{GP} \frac{\operatorname{diag}(m_d, m_u, 0)}{m_u + m_d},\\
    \langle \mathbb{M}^{-1}(\bm{\mathcal{C}}) \rangle &\approx -\langle\bm{c_q^{AP}}\rangle - c_{GP} = -c_{GA}.
\end{split}
\end{equation}
For massive ALPs, we define analogously the shifted operator
\begin{equation}
    \mathbb{M}_a(\bm{A}) = B_0\{\bm{m_q},\bm{A}\} + \frac{m_0^2}{N_f}\langle \bm{A}\rangle \bm{1} - m_a^2 \bm{A},\qquad \mathbb{M}_a(\bm{\lambda_i}) = (m_i^2 -m_a^2)\bm{\lambda_i}\label{eq:operator_shifted}
\end{equation}
and its inverse via spectral decomposition
\begin{equation}
    \mathbb{M}_a^{-1}(\bm{A}) = \frac{1}{2}\sum_i \frac{\langle \bm{A}\bm{\lambda_i}^\dagger\rangle}{m_i^2-m_a^2}\bm{\lambda_i}.
\end{equation}
The elements of the inverse operator applied to an arbitrary matrix $\bm{A}$ are
\begin{equation}\begin{split}
    \big(\mathbb{M}_a^{-1}(\bm{A})\big)_{qq'} &= \frac{(\bm{A})_{qq'}}{B_0 (m_q+m_{q'})-m_a^2},\qquad q\neq q',\\
    \big(\mathbb{M}_a^{-1}(\bm{A})\big)_{qq} &=  \frac{1}{2B_0 m_q-m_a^2} \left[(\bm{A})_{qq} -\frac{\langle \bm{A} (2B_0\bm{m_q}-m_a^2 \bm{1})^{-1}\rangle}{\frac{N_f}{m_0^2} + \langle(2B_0\bm{m_q}-m_a^2 \bm{1})^{-1}\rangle} \right].
\end{split}\label{eq:Ma_inverse_full}
\end{equation}
For $m_a\ll m_\pi$,
\begin{equation}
\mathbb M_a^{-1}(\bm A) = \mathbb M^{-1}(\bm A),
\end{equation}
whereas for $m_a\gg m_{\eta'}$,
\begin{equation}
\mathbb M_a^{-1}(\bm A) = -\frac{\bm A}{m_a^2}.
\label{eq:lims_mixing}
\end{equation}

\section{Diagonalisation of the pseudoscalar meson Lagrangian}
\label{app:mesonmass}

In this appendix, we review the diagonalisation of the pseudoscalar meson Lagrangian 
in the absence of relevant ALP--meson mixing. We work in the basis defined in 
Eq.~\eqref{eq:pseudoscalar}. The masses for the charged mesons are given by
\begin{equation}
    m_{\pi^\pm}^2 = B_0 (m_u + m_d) + \delta_\mathrm{em},\qquad m_{K^\pm}^2 =  B_0 (m_u + m_s) + \delta_\mathrm{em},\qquad m_{K^0}^2 = B_0 (m_d + m_s),
\end{equation}
where $\delta_\mathrm{em}$ represents a small electromagnetic mass splitting given by Dashen's theorem~\cite{Dashen:1969eg}. The mass matrix for the neutal mesons, in the basis $\{\pi_3, \eta_8, \eta_0\}$, is given by
\begin{equation}
    M_{ij}^2 = \begin{pmatrix}
        B_0(m_u+m_d) & \frac{B_0}{\sqrt{3}}(m_u-m_d) & \sqrt{\frac{2}{3}}B_0 (m_u-m_d) \\
        \frac{B_0}{\sqrt{3}}(m_u-m_d) & \frac{B_0}{3}(m_u+m_d+4m_s) & \frac{\sqrt{2}}{3}B_0(m_u + m_d-2m_s)\\
        \sqrt{\frac{2}{3}}B_0 (m_u-m_d) & \frac{\sqrt{2}}{3}B_0(m_u + m_d-2m_s) & \frac{2}{3} B_0 (m_u + m_d + m_s) + m_0^2
    \end{pmatrix}.
\end{equation}
The physical mesons $\{\pi^0, \eta, \eta'\}$ are linear combinations of $\{\pi_3, \eta_8, \eta_0\}$ obtained as the eigenvectors of $M_{ij}^2$, and their masses are the associated eigenvalues.

We fit the parameters $\Theta = \{B_0 m_u, B_0 m_d, B_0 m_s, m_0^2, \delta_\mathrm{em}\}$ using the following inputs:
\begin{itemize}
\item Masses of the pseudoscalar mesons~\cite{ParticleDataGroup:2026aaa}, except for $m_\eta$, which is excluded from the likelihood. An adequate description of $\eta$--$\eta'$ mixing requires at least NLO $\chi$PT, so omitting $m_\eta$ gives the best convergence within our framework.
\item Lattice QCD estimates of the ratios of the quark masses $m_u/m_d$ and $m_s/m_{ud}$~\cite{FlavourLatticeAveragingGroupFLAG:2024oxs}.
\item Lattice QCD estimate of the topological susceptibility~\cite{Vicari:2008jw}.
\end{itemize}
The numerical input values are summarised in Table~\ref{tab:mesonfit_inputs}. 
This procedure is intended as an effective leading-order calibration
of the parameters and neutral-meson flavour matrices used in our
phenomenological analysis, rather than as a statistically complete
determination of the low-energy constants; accordingly, the quoted
posterior uncertainties propagate the adopted input uncertainties
within the leading-order framework and do not include chiral
truncation effects.

\begin{table}[t]
    \centering
    \begin{tabular}{@{}ccc@{}}
    \toprule
    Observable & Input value & Posterior prediction \\
    \midrule
       $m_{\pi^0}$  & $134.9768(5)\,\mathrm{MeV}$ & $134.9768(7)\,\mathrm{MeV}$ \\
        $m_{\pi^\pm}$ & $139.57039(17)\,\mathrm{MeV}$ & $139.5704(3)\,\mathrm{MeV}$\\
        $m_{K^\pm}$ & $493.677(13)\,\mathrm{MeV}$ & $493.62(2)\,\mathrm{MeV}$ \\
        $m_{K^0}$ & $497.611(13)\,\mathrm{MeV}$ & $497.656(18)\,\mathrm{MeV}$\\
        $m_\eta$ & $547.862(17)\,\mathrm{MeV}^{\dagger}$ & $494.10(2)\,\mathrm{MeV}$ \\
        $m_{\eta'}$ & $957.78(6)\,\mathrm{MeV}$ & $957.78(8)\,\mathrm{MeV}$\\
        $m_u/m_d$ & $0.485(19)$ & $0.559(19)$\\
        $m_s/m_{ud}$ & $27.42(12)$ & $25.77(12)$\\
        $\sqrt[4]{\chi_t}$ & $180(3)\,\mathrm{MeV}$ & $175(3)\,\mathrm{MeV}$\\
    \bottomrule
    \end{tabular}
    \caption{Observables used as inputs and their posterior predictions of the parameters used to fit the mesonic quadratic Lagrangian. ${}^\dagger$: The mass of the $\eta$ meson was not included in the likelihood and is shown only for comparison.}
    \label{tab:mesonfit_inputs}
\end{table}

We have implemented a Bayesian fit using the library \texttt{pymc}~\cite{pymc2023}. A likelihood function compares the experimental values $y_i \pm \sigma_i$ with their theoretical expressions $f_i(\Theta)$, assuming Gaussian uncertainties,
\begin{equation}
    L(y|\Theta) = \prod_i \frac{1}{\sqrt{2\pi}\sigma_i}
  \exp\left[-\frac{(y_i-f_i(\Theta))^2}{2\sigma_i^2}\right].
\end{equation}
For the neutral meson masses, the computation of $f_i(\Theta)$ involves obtaining the eigenvalues of $M^2_{ij}$. For $3\times 3$ symmetric matrices, these can be obtained efficiently and stably using the Cardano method~\cite{Smith:1961,Kopp:2006wp}.

Together with the likelihood, the other ingredient of the Bayesian fit is the set of prior distributions. We choose informative priors centred around $B_0 m_u + B_0 m_d \sim m_{\pi^0}^2$, $(B_0 m_u)/(B_0 m_d) \sim m_u /m_d$, $2 (B_0 m_s)/(B_0 m_u + B_0 m_d) \sim m_s/m_{ud}$, and $\delta_\mathrm{em} \sim m_{\pi^\pm}^2 - m_{\pi^0}^2$.

The posterior distribution built from these priors and the likelihood is sampled using the NUTS Markov Chain Monte Carlo algorithm implemented by \texttt{pymc}, with four chains of 2000 samples each. The resulting fit parameters are
\begin{equation}\begin{split}
    B_0 m_u = 6568(9)\,\mathrm{MeV}^2,\quad B_0 m_d = 11740(9)\,\mathrm{MeV}^2,\quad B_0 m_s = 235921(10)\,\mathrm{MeV}^2,\\
    m_0^2 = 0.67124(13)\,\mathrm{GeV}^2,\quad \delta_\mathrm{em} = 1171.9(6)\,\mathrm{MeV}^2, \label{eq:fit_mesonic}
\end{split}\end{equation}
and correlation matrix
\begin{equation}
    \mathrm{Corr} = \begin{pmatrix}
        1 & -1 & 0.107 & -0.019 & 0.972\\
        -1 & 1 & -0.107 & 0.018 & -0.975\\
        0.107 & -0.107 & 1 & -0.091 & 0.106\\
        -0.019 & 0.018  & -0.091 & 1 & 0.005\\
        0.972 & -0.975 & 0.106 & 0.005 & 1
    \end{pmatrix}.
\end{equation}

After diagonalising $M_{ij}^2$ with the numerical results of the fit, the $U(3)$ matrices for the neutral mesons are
\begin{equation}
\begin{split}
    \bm{\lambda_{\pi^0}} &= 0.9998187(13) \bm{\lambda_3} + 0.01654(6) \bm{\lambda_8} + 0.00943(3) \bm{\lambda_0} \\ &= \mathrm{diag}(1.01707(6), -0.98257(6), -0.01140(4)), \\
    \bm{\lambda_\eta} &= -0.01875(7) \bm{\lambda_3} + 0.94125(2) \bm{\lambda_8} + 0.33718(6)\bm{\lambda_0} \\&= \mathrm{diag}(0.79999(8), 0.83749(7), -0.81156(8)),\\
    \bm{\lambda_{\eta'}} &= -0.003302(12)\bm{\lambda_3} - 0.33730(6) \bm{\lambda_8} + 0.94139(2)\bm{\lambda_0} \\&= \mathrm{diag}(0.57061(5), 0.57721(6), 1.15812(5)).
\end{split}\label{eq:matrices_physmesons}
\end{equation}
These numerical results are in reasonable agreement with the approximations commonly used in Refs.~\cite{Aloni:2018vki,Ovchynnikov:2025gpx,Balkin:2025enj},
\begin{equation}
    \bm{\lambda_{\pi^0}} \approx \mathrm{diag}(1, -1, 0),\qquad \bm{\lambda_\eta} \approx \sqrt{\frac{2}{3}}\mathrm{diag}(1, 1, -1),\qquad \bm{\lambda_{\eta'}} \approx \frac{1}{\sqrt{3}}\mathrm{diag}(1, 1, 2).
\end{equation}
The implementation of the fit can be found at \href{https://github.com/alp-aca/mesonfit}{alp-aca/mesonfit}.

\section{Estimate of the independent singlet coupling
\texorpdfstring{$\widetilde S_B$}{S-tilde-B}}
\label{app:stilde_extraction}

This appendix describes an exploratory determination of the independent
singlet coupling $\widetilde S_B$ from the nucleon matrix element of the
topological charge density.
An early attempt to constrain $\widetilde S_B$ was made by Ref.~\cite{Bass:1999is}.  The historical OZI expectation $g_A^{(0)}\simeq0.6$ was inferred from the octet axial charge measured in baryon semileptonic decays, together with the assumption $\Delta s=0$.  The author attributed the difference between this expectation and the contemporary polarised deep inelastic scattering (pDIS) range, $g_A^{(0)}|_{\rm pDIS}\simeq0.2$--$0.35$, entirely to the contribution associated with $\widetilde S_B$ in the singlet Goldberger--Treiman relation.  This yielded an order-of-magnitude estimate $\widetilde{S}_B =-\sqrt{2/3} f_\pi/m_p\,g_{QNN}\sim -0.2$, but relied on a saturation assumption rather than a direct measurement of the topological form factor.  More recently, COMPASS obtained $g_A^{(0)}|_{\rm pDIS}=0.32(7)$~\cite{COMPASS:2016jwv}, while lattice QCD~\cite{Alexandrou:2024ozj} finds $g_A^{(8)}=0.490(20)$ rather than the historical SU(3)-symmetric value $0.585$.  The reduced singlet--octet discrepancy therefore does not support assigning the full historical difference to $\widetilde S_B$ and motivates our extraction from the finite $Q^2$ topological form factor calculated in lattice QCD by Ref.~\cite{Gong:2015iir}.

\subsection{Leading-order topological form factor}
To relate the finite-momentum lattice data to $\widetilde S_B$, we derive
the leading-order $U(3)$ expression for the proton topological form factor
$G_{\widetilde G}(-Q^2)$, with $Q^2=-q^2>0$.  This is the spacelike
elastic counterpart of the form factor introduced for baryon--antibaryon
production in Section~\ref{sec:baryon_GeV_continuation}.  Its matrix
element receives both a direct contribution proportional to
$\widetilde S_B$ and neutral-meson exchange contributions, whose
momentum dependence provides the basis for the extraction below.

We first identify the proton--meson vertices entering the exchange
contributions, and then derive the direct term from the auxiliary
topological density $\omega$.  At vanishing external sources, both
singlet baryon operators contribute to the $\eta_0$ vertex.  Expanding
Eq.~\eqref{eq:baryoncoeff} to linear order in the neutral meson fields gives
\begin{equation}
 \mathcal L_{p\phi}
 =-\frac{1}{2 m_N}\,(\bar p\gamma^\mu\gamma_5p)\,
\left(g_{\pi NN}\partial_\mu \pi_3 + g_{\eta_8 N N}\partial_\mu\eta_8+g_{\eta_0 N N}\partial_\mu\eta_0\right),
 \label{eq:proton_meson_vertex}
\end{equation}
where
\begin{equation}\begin{split}
g_{\pi NN} &= \frac{m_N}{f_\pi}(\Delta u - \Delta d),\\
 g_{\eta_8 N N}&=\frac{m_N}{f_\pi}\frac{\Delta u+\Delta d-2\Delta s}{\sqrt3},
 \\
 g_{\eta_0 N N}&=\frac{m_N}{f_\pi}\sqrt{\frac23}\left(\Delta u+\Delta d+\Delta s+3\widetilde S_B\right).
 \label{eq:x8_x0}
\end{split}\end{equation}

The contact term can be seen before eliminating the auxiliary topological
density.  The terms relevant to a proton matrix element are
\begin{equation}
 \mathcal L_\omega=
 \frac{\omega^2}{2\chi_t}
 +\frac{\omega}{f_\pi}\langle\bm\Phi\rangle
 +\frac{\widetilde S_B}{2\chi_t}
 (\bar p\gamma^\mu\gamma_5p)\partial_\mu\omega.
 \label{eq:omega_auxiliary_lagrangian}
\end{equation}
After integrating the last term by parts, the auxiliary-field equation gives
\begin{equation}
 \omega=-\frac{\chi_t}{f_\pi}\langle\bm\Phi\rangle
 +\frac{\widetilde S_B}{2}\partial_\mu (\bar p\gamma^\mu\gamma_5p).
 \label{eq:omega_eom}
\end{equation}
The first term generates the neutral-meson poles and the second is a direct
nucleon contribution.  Combining them gives the leading-order
$U(3)$ result at spacelike $Q^2=-q^2>0$,
\begin{equation}
 G_{\widetilde G}(-Q^2)=
 -\frac{m_0^2}{6}\,
 \bm v^T\left(Q^2\bm 1+\bm{M}^2\right)^{-1}
 \bm{g_{NN}}(\widetilde S_B)+\widetilde S_B,
 \label{eq:gG_matching}
\end{equation}
with
\begin{equation}
 \bm v=\begin{pmatrix}0\\0\\ \sqrt6\end{pmatrix},
 \qquad
 \bm{g_{NN}}=\frac{m_N}{f_\pi}\begin{pmatrix}
 \Delta u - \Delta d \\[2mm]
 (\Delta u+\Delta d-2\Delta s)/\sqrt3\\[2mm]
 \sqrt{2/3}\,(\Delta u+\Delta d+\Delta s+3\widetilde S_B)
 \end{pmatrix}.
 \label{eq:v_x_definitions}
\end{equation}
Here $m_0^2=2N_f\chi_t/f_\pi^2$, with $N_f=3$.  In the isospin limit the
leading-order mass matrix is
\begin{equation}
 \bm{M}^2=\begin{pmatrix} m_\pi^2 & 0 & 0\\[3mm]
0 &\dfrac{4m_K^2-m_\pi^2}{3}
 &-\dfrac{2\sqrt2}{3}(m_K^2-m_\pi^2)\\[3mm]
 0 & -\dfrac{2\sqrt2}{3}(m_K^2-m_\pi^2)
 &\dfrac{2m_K^2+m_\pi^2}{3}+m_0^2
 \end{pmatrix}.
 \label{eq:M80_LO}
\end{equation}
Since in the isospin limit, as is used in the lattice input used below, the pion decouples, we can restrict our analysis to the $\bm{M_{80}}^2$ block matrix.

As a normalization and sign check, in the flavour-symmetric chiral limit at
$Q^2=0$ Eq.~\eqref{eq:gG_matching} reduces to
\begin{equation}
 G_{\widetilde G}(0)= 
 -\frac{\Delta u+\Delta d+\Delta s+3\widetilde S_B}{3} + \widetilde{S}_B
 =-\frac{\Delta u+\Delta d+\Delta s}{3}.
 \label{eq:chiral_limit_check}
\end{equation}
The cancellation of $\widetilde S_B$ is required by the anomalous Ward
identity.  At finite $Q^2$ and nonzero quark masses the cancellation is
incomplete, which makes the momentum dependence of $G_{\widetilde G}$ sensitive to
$\widetilde S_B$.

As simplifying assumptions, we have discarded the $\mathcal{O}(\alpha_s^2)$ mixing between the topological density and the singlet axial current~\cite{Espriu:1982bw,Gong:2015iir}, and we have described the baryons as pointlike particles: at large $Q^2$ the leading-order coefficients in Eq.~\eqref{eq:x8_x0} should be replaced by momentum-dependent form factors.

\subsection{Lattice and pseudoscalar-spectrum inputs}

We digitized the six values of the separated topological form factor in
Fig.~2 of Ref.~\cite{Gong:2015iir}.  They are listed in
Table~\ref{tab:gG_digitized} and cover
$0.21\lesssim Q^2\lesssim1.08\,\mathrm{GeV}^2$.
\begin{table}[t]
 \centering
 \begin{tabular}{@{}ccccccc@{}}
 \toprule
 $Q^2\,[\mathrm{GeV}^2]$ & 0.207 & 0.400 & 0.581 & 0.753 & 0.917 & 1.074\\
 $g_G(-Q^2) = -G_{\widetilde G}(-Q^2)$              & 0.1039 & 0.0715 & 0.0535 & 0.0478 & 0.0325 & 0.0293\\
 \bottomrule
 \end{tabular}
 \caption{Values digitized from Fig.~2 of Ref.~\cite{Gong:2015iir}.  The
 published plot does not show point-by-point error bars for the separated
 $g_G = -G_{\widetilde G}$ contribution.}
 \label{tab:gG_digitized}
\end{table}
The calculation used a mixed action on the RBC/UKQCD $24^3\times64$ ensemble
with $m_\pi\simeq0.330\,\mathrm{GeV}$ and $a^{-1}=1.73\,\mathrm{GeV}$.
The kaon mass is not quoted by Ref.~\cite{Gong:2015iir}; as an ensemble proxy
we use $a m_K=0.3325(6)$ from Table IX of Ref.~\cite{RBC:2010qam}, giving
$m_K=0.575(10)\,\mathrm{GeV}$.  Because the valence action in
Ref.~\cite{Gong:2015iir} is different, this is not a determination of the
mixed-action kaon mass.

For a more direct estimate of the singlet spectrum, we use Table II of
Ref.~\cite{Christ:2010dd}.  Those ensembles have the same sea action,
coupling, and lattice spacing, but a smaller $16^3\times32$ volume and heavier
light-quark masses.  Following that reference, weighted straight lines are
fitted to $m_\eta^2$ and $m_{\eta'}$ as functions of $m_\pi^2$ and evaluated
at the $24I$ light-quark point $a m_\pi=0.1904(6)$.  The mixing angle is
linearly extrapolated from the two masses at which it is quoted.  This gives
\begin{equation}
 m_\eta=0.599(4)\,\mathrm{GeV},\qquad
 m_{\eta'}=0.986(100)\,\mathrm{GeV},\qquad
 \theta=-9.7(4.0)^\circ.
 \label{eq:christ_extrapolation}
\end{equation}
With the convention
$\bigl(\eta,\eta'\bigr)^T=R(\theta)\bigl(\eta_8,\eta_0\bigr)^T$, the
corresponding central mass matrix is
\begin{equation}
 \bm{M_{80,\mathrm{Christ}}}^2
 =R^T(\theta)\operatorname{diag}(m_\eta^2,m_{\eta'}^2)R(\theta)
 =\begin{pmatrix}
 0.376&-0.102\\
 -0.102&0.955
 \end{pmatrix}\mathrm{GeV}^2.
 \label{eq:christ_matrix}
\end{equation}
At the reweighted physical strange mass Ref.~\cite{RBC:2010qam} gives
$a m_K=0.314(4)$.  The leading-order trace relation then implies
\begin{equation}
 m_0^2=m_\eta^2+m_{\eta'}^2-2m_K^2
 =0.741^{+0.209}_{-0.190}\,\mathrm{GeV}^2,
 \label{eq:christ_m0}
\end{equation}
where the uncertainty is dominated by $m_{\eta'}$.

\subsection{Fit and uncertainty estimate}

For fixed pseudoscalar inputs Eq.~\eqref{eq:gG_matching} is affine in
$\widetilde S_B$.  Writing its value at the $i$th momentum as
$a_i+b_i\widetilde S_B$, the unweighted least-squares estimator is
\begin{equation}
 \widehat{\widetilde S}_B
 =\frac{\sum_i b_i\left[G_{\widetilde G,i}-a_i\right]}
 {\sum_i b_i^2}.
 \label{eq:stilde_estimator}
\end{equation}
We use the axial inputs in Eq.~\eqref{eq:axial_charges_lattice}.  A baseline fit with
$m_\pi=0.330\,\mathrm{GeV}$, $m_K=0.575\,\mathrm{GeV}$, and the value $m_0^2=0.67124\,\mathrm{GeV}^2$ from the fit in Eq.~\eqref{eq:fit_mesonic} gives
$\widetilde S_B=0.0254$.

We use several variations of the fit to estimate the stability of the result:
\begin{itemize}
 \item retaining only the first four, more infrared, points gives
 $\widetilde S_B=0.0207$;
 \item changing to $m_0^2=0.728\,\mathrm{GeV}^2$ gives
 $\widetilde S_B=0.0317$;
 \item using the axial central values quoted by the $\chi$QCD analysis~\cite{Gong:2015iir} gives
 $\widetilde S_B=0.0332$;
 \item subtracting the separately digitized pseudoscalar contribution from
 the Fig.~3 Ward-identity combination gives $\widetilde S_B=0.018$.  The
 Fig.~3 uncertainties belong to the sum and cannot be assigned to $G_{\widetilde G}$ alone.
\end{itemize}

The spectrum of Ref.~\cite{Christ:2010dd} allows two further implementations.  Directly
inserting Eq.~\eqref{eq:christ_matrix}, while using
Eq.~\eqref{eq:christ_m0} for the topological-source normalization, yields
\begin{equation}
 \widetilde S_B=0.0231,
 \qquad 68\%\ \text{Monte Carlo interval:}\ [-0.0003,0.0507],
 \label{eq:christ_direct_result}
\end{equation}
with a residual root-mean-square deviation of $0.0082$.  Alternatively, one
may use~\cite{Christ:2010dd} only to determine $m_0^2$ and then reconstruct the
leading-order matrix in Eq.~\eqref{eq:M80_LO} at the simulated $24I$ sea
strange mass.  This gives
\begin{equation}
 \widetilde S_B=0.0332,
 \qquad 68\%\ \text{Monte Carlo interval:}\ [0.0120,0.0560],
 \label{eq:christ_shifted_result}
\end{equation}
with a residual root-mean-square deviation of $0.0039$.  The intervals in
Eqs.~\eqref{eq:christ_direct_result} and
\eqref{eq:christ_shifted_result} propagate the published statistical errors
independently; they do not include correlations or chiral, finite-volume,
mixed-action, or digitization systematics.  The second implementation is
internally consistent with the leading-order mass matrix, but the shift in
the strange mass is itself a leading-order assumption.

As a phenomenological ``na\"ive'' prescription, we will take the baseline fit result as the central value for $\widetilde{S}_B$, and a symmetric envelope of the alternative fits as an estimate for its uncertainty:
\begin{equation}
    \widetilde{S}_B = 0.025 \pm 0.031\qquad(\text{na\"ive prescription}).
\end{equation}
Additionally, we also consider a more ``conservative'' prescription, due to the unaccounted effects related to $\mathcal{O}(\alpha_s^2)$ mixing and deviations from pointlike baryons, increasing the error by a factor two,
\begin{equation}
    \widetilde{S}_B = 0.025 \pm 0.062\qquad(\text{conservative prescription}).
\end{equation}
These ranges should not be interpreted as a lattice uncertainty.

\section{Implementation in \texttt{ALP-aca}}
\label{app:alpaca}

Version 1.2 of \texttt{ALP-aca}~\cite{Alda:2025nsz} implements the
ALP--baryon interactions derived in this work.
The new function \texttt{alpaca.effective\_coupling} implements the effective couplings to baryons. For example, for an ALP with mass $m_a = 1.2\,\mathrm{GeV}$ and parton-level couplings contained in the object \texttt{cG}\footnote{Note that the definition of the quark couplings in \texttt{ALP-aca} differs to the one used in this work by a factor of 2. We refer the reader to the manual~\cite{Alda:2025nsz} for the definitions used in \texttt{ALP-aca}.}, the effective coupling $c_{\Sigma^+\Sigma^+}^A$ is calculated as
\begin{verbatim}
c_SigmaSigma = alpaca.effective_coupling(1.2, cG, 'Sigma+ Sigma+', 'A')
\end{verbatim}

The calculation of the effective couplings to baryons involves three optional parameters, detailed in Table~\ref{tab:alpaca_pars}, controlling the two main sources of model-dependence in our analysis: \texttt{sbtilde} determines the value of $\widetilde{S}_B$; and \texttt{mu\_pQCD\_baryons} and \texttt{formfactors\_baryons\_spline} the position and shape of the form factors modelling the heavy ALP continuation.

The functions \texttt{alpaca.decay\_width} and \texttt{alpaca.branching\_ratio} have been extended to include ALP decays into baryons, and processes where the ALP is produced in FCNC baryon decays. All functions also accept the optional parameters in Table~\ref{tab:alpaca_pars} and propagate them to the calculation of the effective baryon couplings.

The database of experimental measurements includes the measurements of $\Sigma^+ \to p + \mathrm{inv}$~\cite{BESIII:2023utd} and $\Xi^0 \to \Lambda + \mathrm{inv}$~\cite{BESIII:2026ztz} at BESIII and $\Sigma^+ \to p \mu^+ \mu^-$ at LHCb~\cite{LHCb:2017rdd,LHCb:2025evf}.

\begin{table}[t]
    \centering
    \begin{tabular}{@{}ccl@{}}
    \toprule
      Argument   & Parameter & Comment  \\
    \midrule
       \texttt{sbtilde}  &  $\widetilde{S}_B$ & Default \texttt{sbtilde=0.025}. \\
       \texttt{mu\_pQCD\_baryons} & $\mu_\star$ & Default \texttt{mu\_pQCD\_baryons=1.9}.\\
       \multirow{2}{*}{\texttt{formfactors\_baryons\_spline}} & \multirow{2}{*}{$\mathcal{R}(q^2)$} & Default (\texttt{False}): piecewise function in Eq.~\eqref{eq:baryon_continuation_profile}, \\
       & & \texttt{True}: Cubic spline interpolation.\\
    \bottomrule
    \end{tabular}
    \caption{Optional parameters entering the calculation of baryon effective couplings in \texttt{ALP-aca} version 1.2.}
    \label{tab:alpaca_pars}
\end{table}

\renewcommand{\bibsection}{\section*{References}}
\bibliography{biblio}

\end{document}